\documentclass[review]{elsarticle}

\usepackage{lineno,hyperref}
\usepackage{amsfonts}
\usepackage{graphicx}
\usepackage{subfig}

\modulolinenumbers[5]

\begin{document}

\begin{frontmatter}

\title{Bearing fault diagnosis under varying working condition based on domain adaptation}

\author[mymainaddress,mysecondaryaddress]{Bo Zhang}

\author[mymainaddress]{Wei Li\corref{mycorrespondingauthor}}
\cortext[mycorrespondingauthor]{Corresponding author. E-mail addresses: \emph{liwei\_cmee@163.com} }

\author[mymainaddress]{Zhe Tong}

\author[mymainaddress]{Meng Zhang}

\address[mymainaddress]{School of Mechatronic Engineering, China University of Mining and Technology Xuzhou 221116, People’s Republic of China}
\address[mysecondaryaddress]{School of Computer Science And Technology, China University of Mining and Technology Xuzhou 221116, People’s Republic of China}

\begin{abstract}

Traditional intelligent fault diagnosis of rolling bearings work well only under a common assumption that the labeled training data (source domain) and unlabeled testing data (target domain) are drawn from the same distribution. However, many real recognitions of bearing faults show disobedience of this assumption, especially when the working condition varies. In this case, the labeled data obtained in one working condition may not follow the same distribution in another different working condition. When the distribution changes, most fault diagnosis models need to be rebuilt from scratch using newly recollected labeled training data. However, it is expensive or impossible to annotate huge amount of training data to rebuild such new model. Meanwhile, large amounts of labeled training data have not been fully utilized yet, which is apparently a waste of resources. As one of the important research directions of transfer learning, domain adaptation (DA) typically aims at minimizing the differences between distributions of different domains in order to minimize the cross-domain prediction error by taking full advantage of information coming from both source and target domains. In this paper, we present one of the first studies on unsupervised DA in the field of fault diagnosis of rolling bearings under varying working conditions and a novel diagnosis strategy based on unsupervised DA using subspace alignment (SA) is proposed. After processed by unsupervised DA with SA, the distributions of training data and testing data become close and the classifier trained on training data can be used to classify the testing data. Experimental results on the 60 domain adaptation diagnosis problems under varying working condition in Case Western Reserve benchmark data and 12 domain adaptation diagnosis problems under varying working conditions in our new data are given to demonstrate the effectiveness of the proposed method. The proposed methods can effectively distinguish not only bearing faults categories but also fault severities.

\end{abstract}


\end{frontmatter}

\section{Introduction}

Bearings are the most commonly used components in rotating machinery, and bearing faults may result in significant breakdowns, and even casualties \cite{Albrecht1987,Jardine2006}. It's important to diagnose bearings and diagnose methods are mainly based on vibration analysis. In recent years, artificial intelligence techniques have been introduced and reported in fault diagnosis of bearings to realize massive data analysis and automatical fault diagnosis, such as support vector machine (SVM), genetic programming, artificial neural networks (ANNs), etc. \cite{Widodo2007,ZHANG2005271,Samanta2003}. Muruganatham et al. \cite{Muruganatham2013} developed a method that singular values in singular analysis were used as extracted features and an artificial neural network (ANN) was applied into fault diagnosis. Jia et al. \cite{JIA2016303} utilized deep neural networks (DNNs) to classify the bearing health conditions and taken Fourier amplitudes from fast Fourier transformation as the input of DNNs. Moreover, various time-domain and time-frequency-domain parameters were extracted by  Jin et al. \cite{Jin2014} and Trace Ratio LDA was utilized as the feature selection method. Various statistical features were extracted by Sugumaran et al. \cite{Sugumaran2007}, which form a feature set, and the decision tree was used to generate the rules automatically to select features from the feature set, for fuzzy classifier. Trendafilova et al. \cite{Trendafilova2010} introduced features extracted by wavelet as the input of PCA and the principal components generated by PCA are used as the input of a nearest neighbor classifier. Zhang et al. \cite{Zhang2013} presented a procedure based on ensemble empirical mode decomposition (EEMD) and optimized support vector machine (SVM) for multi-fault diagnosis of rolling element bearings. Zhou et al. \cite{Zhou2016} extracted features based on shift-invariant dictionary learning (SIDL) and hidden Markov model (HMM) was addressed for machinery fault diagnosis. In addition, the  statistical features were extracted by Li et al. \cite{Li2015} based on the central limit theory and SVM and ANNs were used to classify faults of bearing respectively.

Intelligent fault diagnosis actually has been a great success. However, most of the above proposed methods are only applicable to the situation that the data used to train classifier and the data for testing are under the same working condition, which means that these proposed methods work well only under a common assumption: the training and test data is drawn from the same feature space and the same distribution. In fact, vibration signals used for diagnosis usually show disobedience of the above assumption. In the running process of rotating machinery, because of complicated working conditions and dynamic signal acquisition environment, the distributions of fault data under varying working condition are not consistent. Many diagnosis methods have poor domain adaptation ability. For solving this problem, Li et al. \cite{Li2016} try to extract the features which are insensitive to the changes of working condition, by processing the spectrum images, generated by fast Fourier transformation, with two-dimensional principal component analysis (2DPCA). Unfortunately, the problem has not been solved completely. The applicable working conditions of above method are not enough and the fault classification performance still has the room for improvement. Domain adaptation in transferring learning is to solve this kind of problem. So, we turn to domain adaptation for a solution.

Domain adaption (DA) has aroused large amounts of interest and research in the literatures. DA can be considered as a special setting of transfer learning which aims at transferring shared knowledge across different but related tasks or domains \cite{5288526}, which typically aims at taking full advantage of information coming from both source and target domains during the learning process to adapt automatically. DA is applied in many areas in recent years, such as sentiment analysis \cite{DBLP:journals/jmlr/BlitzerKF11,Vuong:2014:TLE:2682648.2682834,7817047}, visual Object Recognition \cite{6909714,7815350}, handwriting recognition, and cross-Domain WiFi Localization \cite{4444176}.  Blitzer et al. investigated domain adaptation for sentiment classifiers, focusing on textual domain adaptation for online reviews of different types of products, which is the prior work on Domain adaptation \cite{DBLP:journals/jmlr/BlitzerKF11,DBLP:conf/emnlp/BlitzerMP06,DBLP:conf/acl/BlitzerDP07}. Domain adaptation was also applied into emotional polarity classification, such as  reviews of different consumer products, services, and forensic analysis, where distributions of the new testing set and training set are also different. VUONG et al. \cite{Vuong:2014:TLE:2682648.2682834} proposed an adaptation transfer learning approach to utilize the labeled data available for solving the related but different problems. Bravo-Marquez et al. \cite{7817047} focused on sentiment classification of tweets and proposed a simple model for transferring sentiment labels from words to tweets and vice versa by representing both tweets and words using feature vectors residing in the same feature space to avoid the annotation of words or tweets based on polarity classes. Moreover, domain adaptation was also used solve the problem, in indoor WiFi localization that when we predict a mobile user's location based on the received WiFi signals on the mobile device, the distribution of WiFi signal strength constantly changes due to the change of indoor environment.  Yang et al. \cite{4444176} proposed a dimensionality reduction method for DA, which learned a low-dimensional latent feature space where the distributions between the source domain data and the target domain data are the same or close to each other, to address indoor WiFi localization problems. Xu et al. \cite{7855802} presented a metric transfer learning framework (MTLF) to bridge the difference in distributions between the source domain and the target domain. The aforementioned applications prove that DA a promising tool in dealing with the problem that distributions of source domain and target domain are different. However, it attracts few attentions in the field of fault diagnosis.

Through domain adaptation, this paper proposed a novel intelligent diagnosis method to overcome the problem that the distributions of varying conditions are different and the classifier trained under one condition can not be used to classify faults under other condition, by applying unsupervised domain adaption with subspace alignment. This method can take full advantage of information coming from training data and testing data. There are two different scenarios: (1) the unsupervised setting where the target domain data are fully unlabeled; (2) the semi-supervised case where a few labels are provided for the target domain. We focus on the unsupervised domain adaptation setting and assume that $D_T$ are unlabeled, since it does not require any labeling information from the target domain which is well suited to fault diagnosis of bearings. The rest of this paper is organized as follows. Section \ref{sec:method} presents fault diagnosis method based on unsupervised domain adaption with subspace alignment, including subspace generation with FFT, unsupervised domain adaption with subspace alignment,  classification strategy and domain discrepancy analysis. Section \ref{sec:Experimental} presents the experimental analysis and discussion. The conclusion are given in Section \ref{sec:conclusion}.

\section{Fault diagnosis based on Unsupervised domain adaption with subspace alignment}
\label{sec:method}

\begin{figure*}[!ht]
\centering
\includegraphics[width=\textwidth]{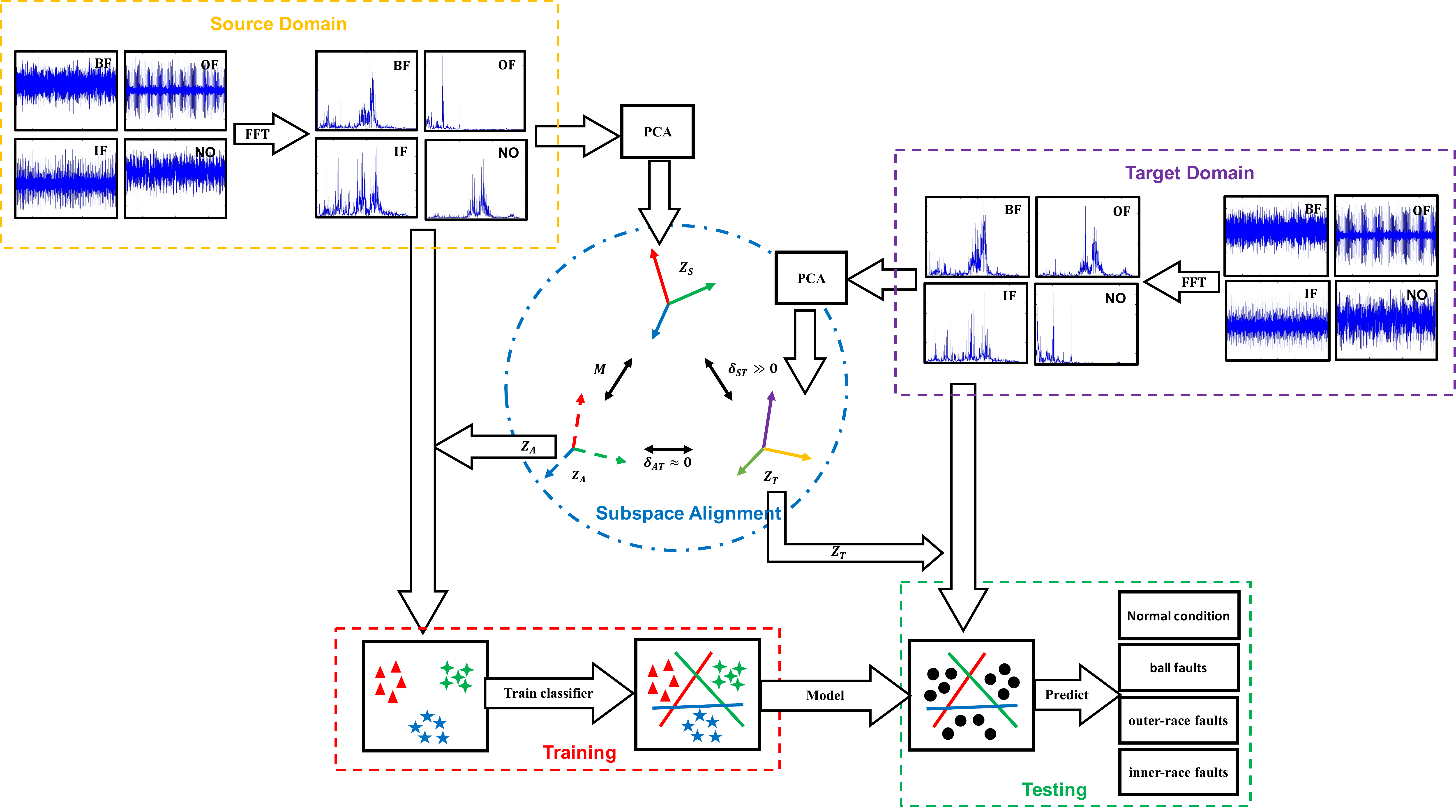}
\caption{The framework of bearing fault diagnosis based on unsupervised DA with SA. The source domain is represented by the source subspace
$Z_S$ and the target domain by target subspace $Z_T$.}
\label{Fig2}
\end{figure*}

In this paper, we study the problem of domain adaptation for bearing fault diagnosis. We focus on the setting that there are only one source and one target domain sharing the same feature space and the same set of fault types. Let $D_S=\{(x_{S_1},y_{S_1}),...,(x_{S_{n_1}},y_{S_{n_1}})\}$, where $x_{S_i} \in \mathcal{X}$ is the input and $y_{S_i} \in \mathcal{Y}$ is the corresponding output. Similarly, let the target domain data be $D_T=\{(x_{T_1},y_{T_1}),...,(x_{T_{n_2}},y_{T_{n_2}})\}$, where the input $x_{T_i} \in \mathcal{X}$. Let $P(X_S)$ and $Q(X_T)$ be the marginal distributions of $X_S=\{x_{S_i}\}$ and $X_T=\{x_{T_i}\}$ from the source and target domains, respectively. In general, $P(X_S)$ and $Q(X_T)$ can be different. Our task is to predict the labels $y_{T_i}$s corresponding to input $x_{T_i}$s in the target domain. The key assumption in most domain adaption methods is that $P(X_S) \neq Q(X_T)$, but $P(Y_S \vert X_S)=Q(Y_T \vert X_T)$. Under these assumptions, we study how to predict the fault types of bearing accurately in the target domain with a different data distribution. In this section, we present our bearing fault diagnosis based on unsupervised domain adaption with subspace alignment (unsupervised DA with SA).  The framework of this procedure is shown in Figure \ref{Fig2}.

\subsection{Subspace generation with FFT}

Considering the setting that there are only one source and one target domain sharing the same feature space and the same set of fault types, we have a set $X_S=\{x_{S_i}\}$ of labeled data (resp. a set $X_T=\{x_{T_i}\}$ of unlabeled data), both lying in the same feature space $\mathcal{X}$ extracted from vibration signals and drawn i.i.d according to a fixed but unknown source (resp. target) distribution $P(X_S)$ (resp. $Q(X_T)$) under different load conditions and rotating speeds.

\begin{figure}[!ht]
\centering
\includegraphics[width=0.45\textwidth]{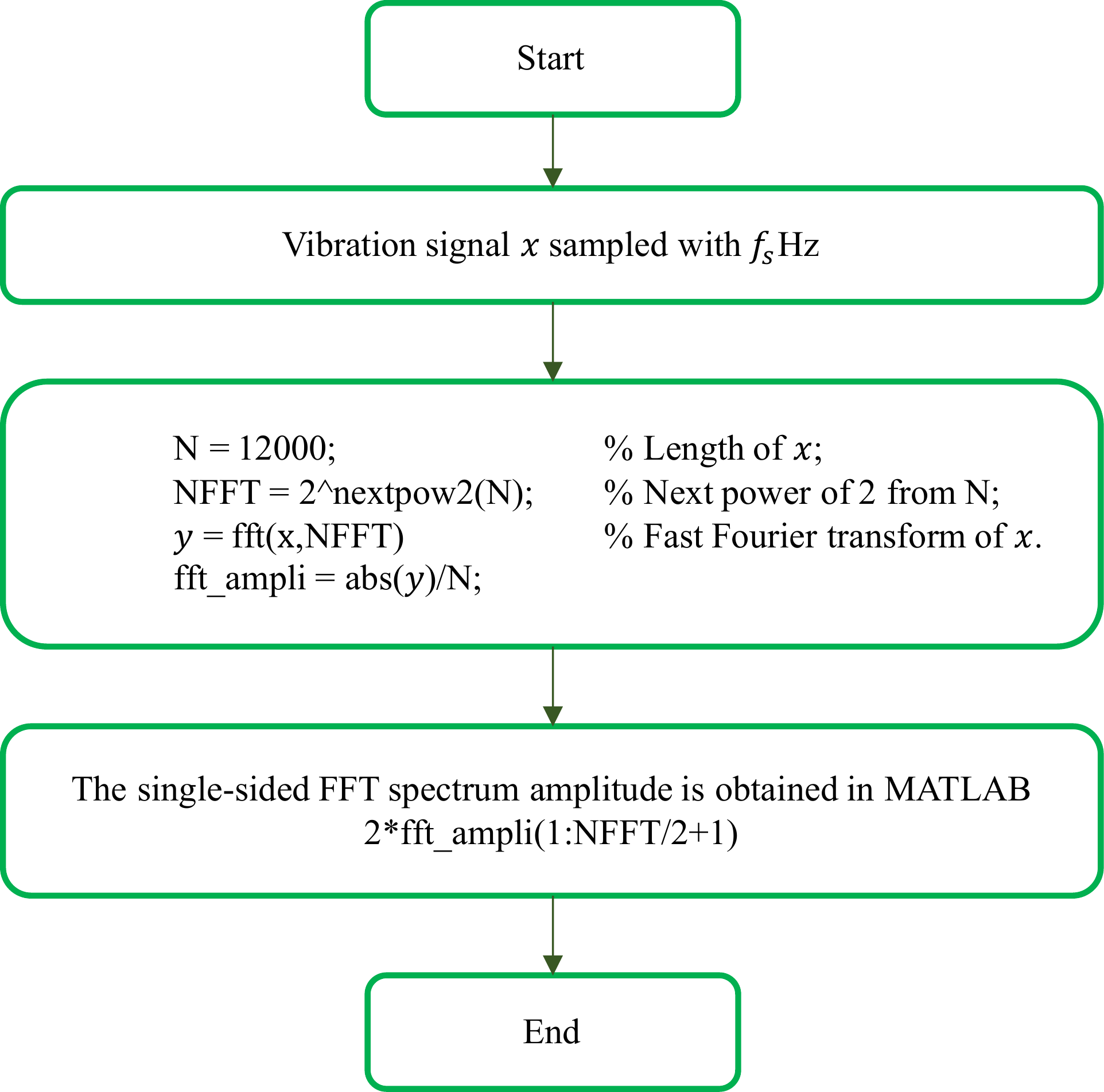}
\caption{Flowchart of computing FFT spectrum amplitudes in MATLAB.}
\label{Fig3_1}
\end{figure}

In this study, we take the fast Fourier transform (FFT) spectrum amplitudes of vibration signals as features which is the most widely used approach of bearing defect detection. The flowchart of computing FFT spectrum amplitudes of vibration signal $x$ sampled with $f_s$ Hertz using $N$ sampling points in MATLAB is detailed illustrated in Figure \ref{Fig3_1}.

In order to diagnose the unlabeled testing data more accurately, we need to get more robust representations of the training data and learn the shift between these two domains. According to Ref. \cite{6751479}, we transform every source and target data to a $D$-dimensional $z$-normalized vector (i.e. of zero mean and unit standard deviation) firstly. Note that $z$-normalized is an important step in most of the subspace-based domain adaption methods such as GFK \cite{6247911} and GFS \cite{6126344}. Then, using PCA, we select for each domain the $d$ eigenvectors corresponding to the $d$ largest eigenvalues. These eigenvectors are used as basis vectors of the source and target subspaces, respectively denoted by $Z_S$ and $Z_T$ ($Z_S, Z_T \in \mathbb{R}^{D \times d}$). Note that $Z^{'}_S$ and $Z^{'}_T$ are orthonormal matrix. Thus, $Z^{'}_SZ_S=I_d$ AND $Z^{'}_TZ_T=I_d$, where $I_d$ is the identity matrix of size $d$.

In the above procedure of generating subspace, we need to tune only one hyper parameter $d$ that controls the dimensionality of subspaces $Z_S$ and $Z_T$ ($Z_S, Z_T \in \mathbb{R}^{D \times d}$). To address this, we choose to leverage the theoretical bound deduced by Fernando et al. \cite{6751479} to select the maximum dimensionality $d_{max}$ to guide the selection process.

We use the training data from the source domain to generate the source subspace expanded by $Z_S$, and data from the target domain to generate the target subspace by $Z_T$. With a slight abuse of notations, we refer $Z_S$ and $Z_T$ as subspaces, where we actually refer to the basis vectors of the subspace.

\subsection{Unsupervised domain adaption with subspace alignment}

As illustrated by recent results \cite{6751479,6247911,6126344}, unsupervised DA with SA approaches seem to be effective to tackle unsupervised domain adaptation problems.

In general, many subspace based domain adaption strategies share the same principle: first they compute a domain specific $d$-dimensional subspace for the source data domain and another one for the target data domain, which is typically composed of $d$ eigenvectors induced by a PCA. Then they project source and target data into intermediate subspaces along the shortest geodesic path connecting the two $d$-dimensional subspaces on the Grassmann manifold \cite{6247911,6126344}. The characteristic is to look for the intermediate subspaces connecting the source domain $D_S$ and target domain $D_T$. Comparing with them, the main difference of subspace alignment algorithm proposed in Ref. \cite{6751479} is to align the two subspaces directly instead of hunting for the intermediate subspaces.

The shift between the two domains can be described by the following Bregman matrix divergence:
\begin{equation}
  {\delta_{ST}=||Z_S-Z_T||^2_F}
\end{equation}
where $||\cdot||^2_F$ is the Frobenius norm.

The shift $\delta_{ST}$ is quite large at the beginning. In order to correct this shift, the source subspace $Z_S$ is aligned with the target ones by a transformation matrix $M$, which defines a movement that pushes $Z_S$ close to target subspace $Z_T$ conceptually. The idea behind this method is illustrated in Figure \ref{Fig2}. The resulting aligned subspace is denoted by $Z_A = Z_SM$. At this time, $Z_A$ looks similar to $Z_T$, i.e. ${\delta_{AT}=||Z_SM-Z_T||^2_F} \approx 0$.

Fernando et al. \cite{6751479} chooses to minimize the following Bregman matrix divergence:
\begin{equation}
  {F(M)=||Z_SM-Z_T||^2_F}
\end{equation}
\begin{equation}
  {M^{*}=argmin_M(F(M))}
\end{equation}

In above formulation, $Z_S$ and $Z_T$ is intrinsically regularized and the Frobenius norm is invariant to orthonormal operations, above equation can be depicted by
\begin{equation}
  {F(M)=||Z^{'}_SZ_SM-Z^{'}_SZ_T||^2_F=||M-Z^{'}_SZ_T||^2_F}
\end{equation}
where $'$ is the transpose operation. Under this paradigm, a closed-form optimal $M^{*}$ can be obtained as $M^{*}=Z^{'}_SZ_T$, and $Z_A = Z_SZ^{'}_SZ_T$.

\subsection{Classification strategy}

Finally, labeled instances $X_S$ from the source domain are projected by $Z_A$ and are used to train the classification model at the training stage. At the test stage, unlabeled instances $X_T$ from the target domain are projected by $Z_T$ and are predicted with the learned model. The more appropriate an alignment is, the better classification results should achieve.

In order to compare labeled source instances $X_S$ with unlabeled target instances $X_T$, a similarity function $Sim(X_S,X_T)$ is defined as follows:

\begin{equation}
  {Sim(X_S,X_T)=X_SZ_AZ^{'}_TX^{'}_T}
\end{equation}

$Sim(X_S,X_T)$ can be directly used to perform a $k$-nearest neighbor classification task. However, since $Sim(X_S,X_T)$ is not positive semi-definite matrix we can not make use of it to learn a SVM directly. By using the software LIBSVM \cite{CC01a}, we apply $X_SZ_AZ^{'}_TX{'}_S$ as the precomputed kernel matrix to train a SVM model and predict the target values of the test data by the kernel matrix $X_SZ_AZ^{'}_TX{'}_T$. In order to prevent the overfitting problem, the cross-validation procedure is used to identify best parameter $C$ by exponentially growing sequences (for example, $C = 10^{-3}, 10^{-2}, 10^{-1}, 10^{0}, 10^{1}, 10^{2}, 10^{3}, 10^{4}$).

\subsection{Domain discrepancy analysis}

According to our experiment, we find that unsupervised DA with SA achieves higher classification accuracy than other non-domain adapting approaches. In this section, we attempt to explain why it works.

Ben-David et al. \cite{Ben-David2010,Ben-David:2006:ARD:2976456.2976474} established the generalization error on the target error $\epsilon_T(h)$ which depends on the source error $\epsilon_S(h)$ and a divergence measure $d_{H{\Delta}H}(P(X_S),Q(X_T))$, called the $H{\Delta}H$ divergence, between the source and target distributions $P(X_S)$ and $Q(X_T)$:

\begin{equation}
  {\epsilon_T(h)=\epsilon_S(h)+d_{H{\Delta}H}(P(X_S),Q(X_T))+\lambda}
  \label{eqdomaindiscrepancy}
\end{equation}
where $h$ is a learned hypothesis, and $\lambda$ the error of the ideal joint hypothesis on $D_S$ and $D_T$, which is supposed to be a negligible term in the case of DA.

Eq. \ref{eqdomaindiscrepancy} tells us that to adapt well, one has to learn a hypothesis $h$ which works well on $D_S$ while reducing the  $H{\Delta}H$ divergence  between $P(X_S)$ and $Q(X_T)$. Estimating $H{\Delta}H$ divergence for a finite sample is exactly the problem of minimizing the empirical risk of a linear classifier $\hat{h}$ that discriminates between instances drawn from $D_S$ and instances drawn from $D_T$, respectively pseudo-labeled with 0 and 1.

More specifically, it involves the following steps:

\begin{enumerate}
  \item Pseudo-labeling the source and target instances with 0 and 1, respectively.
  \item Randomly sampling two sets of instances as the training and testing set.
  \item Learning a linear classifier $\hat{h}$ on the training set and verifying its performance on the testing set.
  \item Estimating the distance as $\hat{d}_{H{\Delta}H}(P(X_S),Q(X_T))=2(1-2err(\hat{h}))$, where $err(\hat{h})$ is the test error.
\end{enumerate}

It's obvious that if two domains perfectly overlap with each other, $err(\hat{h}) \approx 0.5$, and $\hat{d}_{H{\Delta}H}(P(X_S),Q(X_T)) \approx 0$. On the contrary, if two domains are completely distinct from each other, $err(\hat{h}) \approx 0$, and $\hat{d}_{H{\Delta}H}(P(X_S),Q(X_T)) \approx 2$. Therefore, $\hat{d}_{H{\Delta}H}(P(X_S),Q(X_T)) \in [0,2]$. The lower the value is, the smaller two domains divergence.

Now we can empirically evaluate the domain divergence of unsupervised DA with SA. In section \ref{subsection:case1_result}, we will take a typical case of adapting FFT amplitudes from one domain to another domain as an example to visualize the feature representation distribution of two domains after our unsupervised DA with SA. According to the visualization of data distributions, we observe that, unsupervised DA with SA do have pushed the same class of different domains close to each other.

\section{Experimental analysis}
\label{sec:Experimental}

In this section, two diagnosis cases on two databases are used to demonstrate the effectiveness of the proposed method, respectively. Database A is provided by the bearing data centre of Case Western Reserve University (CWRU) \cite{Case-Western-Reserve-University} and database B is obtained by the machinery fault simulator in prof Li lab.    

\subsection{Case 1: Fault diagnosis based on Database A}

\subsubsection{Experimental setup and database preparation}

\begin{figure}
  \centering
  \includegraphics[width=0.6\textwidth]{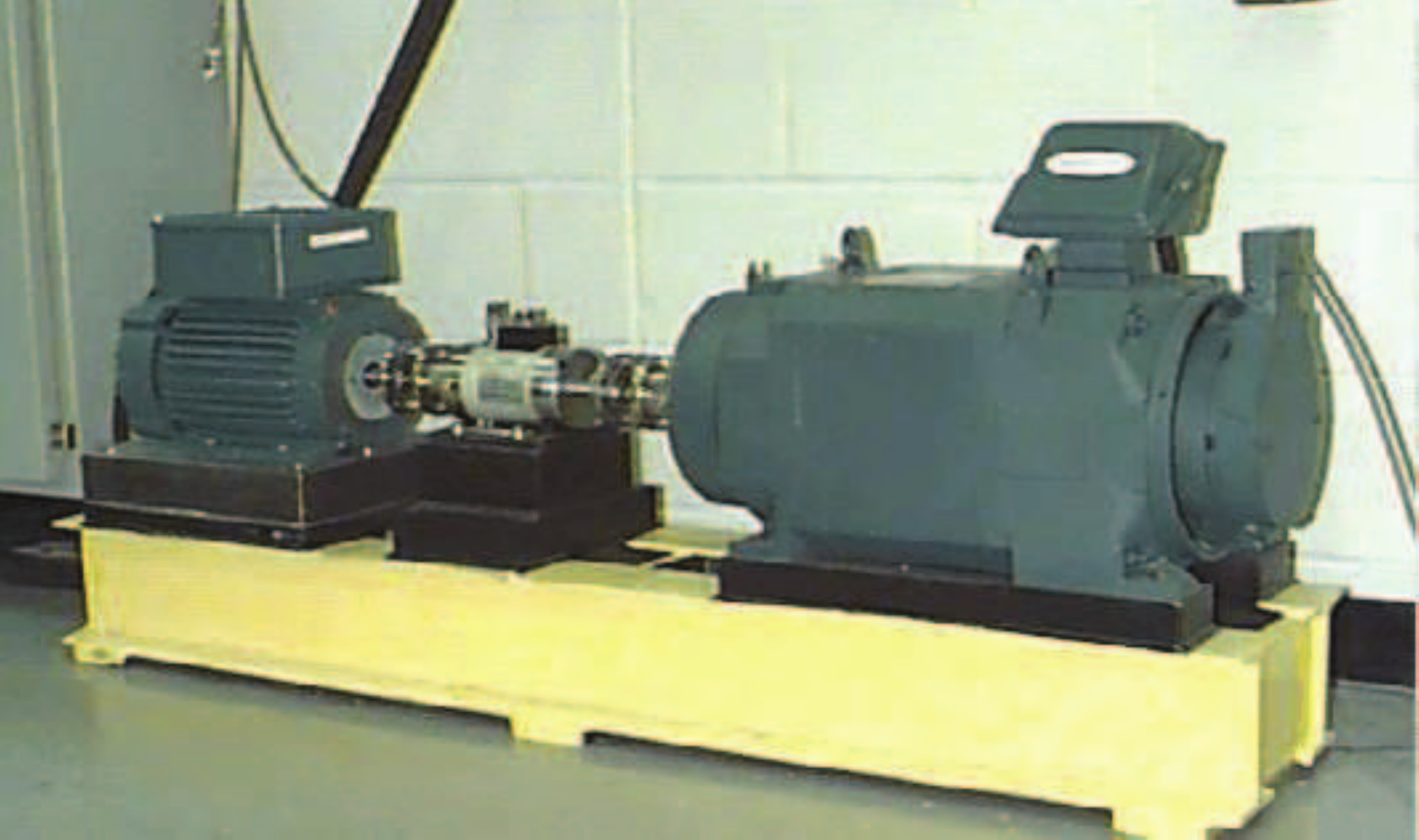}
  \caption{Bearing test rig of Case Western Reserve University Data Center}
\label{Fig_CASE2_setup}
\end{figure}

The database A is provided by Case Western Reserve University (CWRU) Bearing Data Center \cite{Case-Western-Reserve-University}. The test-bed shown in Figure \ref{Fig_CASE2_setup} is composed of a driving motor, a 2 hp motor for loading, a torque sensor/encoder, a power meter, accelerometers and electronic control unit. The test bearings locate in the motor shaft. Subjected to electrosparking, inner-race faults (IF), outer-race faults (OF) and ball fault (BF) with different sizes (0.007in, 0.014in, 0.021in and 0.028in) are introduced into the drive-end bearing of motor. The vibration signals are sampled by the accelerometers attached to the rack with magnetic bases under the sampling frequency of 12kHz. The experimental scheme simulates four load conditions with different rotating speeds, i.e. Load0 = 0hp/1797rpm, Load1 = 1hp/1772rpm, Load2 = 2hp/1750rpm and Load3 = 3hp/1730rpm. The vibration signals of normal bearings (NO) under different load conditions were also gathered.

\begin{table}[htbp]
\begin{center}
  \caption{Description of the experiment datasets on database A}
  \label{Table_case_test_set}
  \begin{tabular}{lccccl}
    \hline
    No. & Datasets & Load(hp) & Fault type & Fault size & Sample size\\
    \hline
    1 &Load0\_0.007 & Load0 & IF,BF,OF,NO & 0.007in & 400  \\
    2 &Load1\_0.007 & Load1 & IF,BF,OF,NO  & 0.007in & 400 \\
    3 &Load2\_0.007 & Load2 & IF,BF,OF,NO & 0.007in & 400 \\
    4 &Load3\_0.007 & Load3 & IF,BF,OF,NO & 0.007in & 400 \\
    5 &Load0\_0.014 & Load0 & IF,BF,OF,NO & 0.014in & 400 \\
    6 &Load1\_0.014 & Load1 & IF,BF,OF,NO & 0.014in & 400 \\
    7 &Load2\_0.014 & Load2 & IF,BF,OF,NO & 0.014in & 400 \\
    8 &Load3\_0.014 & Load3 & IF,BF,OF,NO & 0.014in & 400 \\
    9 &Load0\_0.021 & Load0 & IF,BF,OF,NO & 0.021in & 400 \\
    10 &Load1\_0.021 & Load1 & IF,BF,OF,NO & 0.021in & 400 \\
    11 &Load2\_0.021 & Load2 & IF,BF,OF,NO & 0.021in & 400 \\
    12 &Load3\_0.021 & Load3 & IF,BF,OF,NO & 0.021in & 400 \\
    13 &Load0\_0.028 & Load0 & IF,BF,OF,NO & 0.028in & 400 \\
    14 &Load1\_0.028 & Load1 & IF,BF,OF,NO & 0.028in & 400 \\
    15 &Load2\_0.028 & Load2 & IF,BF,NO & 0.028in & 400 \\
    16 &Load3\_0.028 & Load3 & IF,BF,NO & 0.028in & 400 \\
    \hline
  \end{tabular}
  \end{center}
\end{table}

Four health conditions data (IF, OF, BF, NO) under one load condition with one certain fault size compose one load condition data (namely in Table \ref{Table_case_test_set}). There are four datasets under every fault size and there are 16 datasets totally. In this experiment, we process data sets in a domain adaption setting firstly. The basic idea is to utilize the hierarchy of the data sets. With regard to the datasets under fault size of 0.007in, we can group three domain adaption diagnosis problems using the dataset with certain working condition (e.g. Load0\_0.007) as the testing data and the other three datasets (i.e. Load1\_0.007, Load2\_0.007, Load3\_0.007) as training data respectively. In this case, distributions between the training and testing data may be very different but related. After such pretreatment, four datasets under different working conditions with each fault size can be grouped into 12 domain adaptation diagnosis problems and there are 48 domain adaptation diagnosis problems totally for CWRN benchmark data. There are 100 signals for each health condition and each signal contain 12000 data points. The fast Fourier transformation (FFT) is implemented on each signal and 8193 Fourier amplitudes, which compose one sample, are generated, for convenience of calculations. So each dataset contains 400 samples. In each test, with one certain fault size, one dataset is selected as the target domain, and the other three datasets are chosen as the source domains respectively. For avoiding the random factors, each test was conducted twenty times, and the final accuracy take the average accuracy.

For comparison, contrast experiments: Baseline 1, Baseline 2, SVM with no adaptation (SVM NA), NN with adaptation (NN SA) and SVM with adaptation (SVM SA) are conducted simultaneously.

\begin{itemize}
  \item Baseline 1: 1-Nearest-Neighbor classifier with no adaptation and no projection is made, i.e. we use the original input space without learning a new representation.
  \item Baseline 2: 1-Nearest-Neighbor classifier with no adaptation and a new representation is learned by projecting both source and target data to the PCA subspace $Z_{ST}$ built from both source and target domains. According to the feature dimensionality reduction criterion in Ref. \cite{Abdi2010}, the so called contribution of selected components with 20\%, 40\%, 60\%, 80\%, 90\% and 100\% are firstly performed to determine the reduced dimension. Thereafter, dimension reduction with 90\% is designated in our research, as this was the case in which the average classification rate was the highest.
  \item SVM NA: By using the software LIBSVM \cite{CC01a}, a linear SVM classifier with no adaptation and no projection is made. The cross-validation procedure is used to identify best parameter, cost ($C$) by exponentially growing sequences (for example, $C = 10^{-3}, 10^{-2}, 10^{-1}, 10^{0}, 10^{1}, 10^{2}, 10^{3}, 10^{4}$).
\end{itemize}

\subsubsection{Diagnosis results of the proposed method}
\label{subsection:case1_result}

\begin{figure*}[!ht]
\centering
  \subfloat[]{\includegraphics[width=0.5\textwidth]{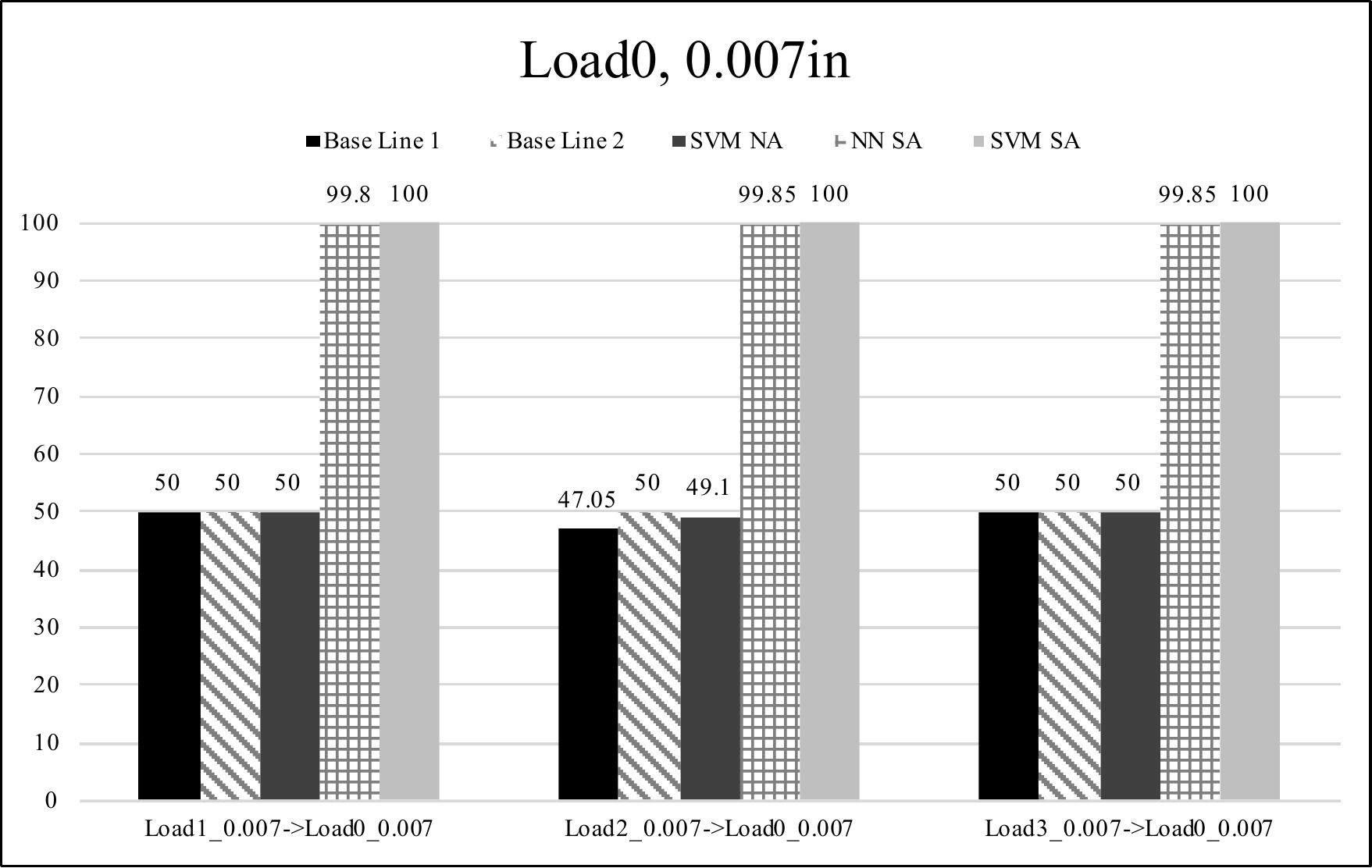}
  \label{fig_0_7_case}}
  \subfloat[]{\includegraphics[width=0.5\textwidth]{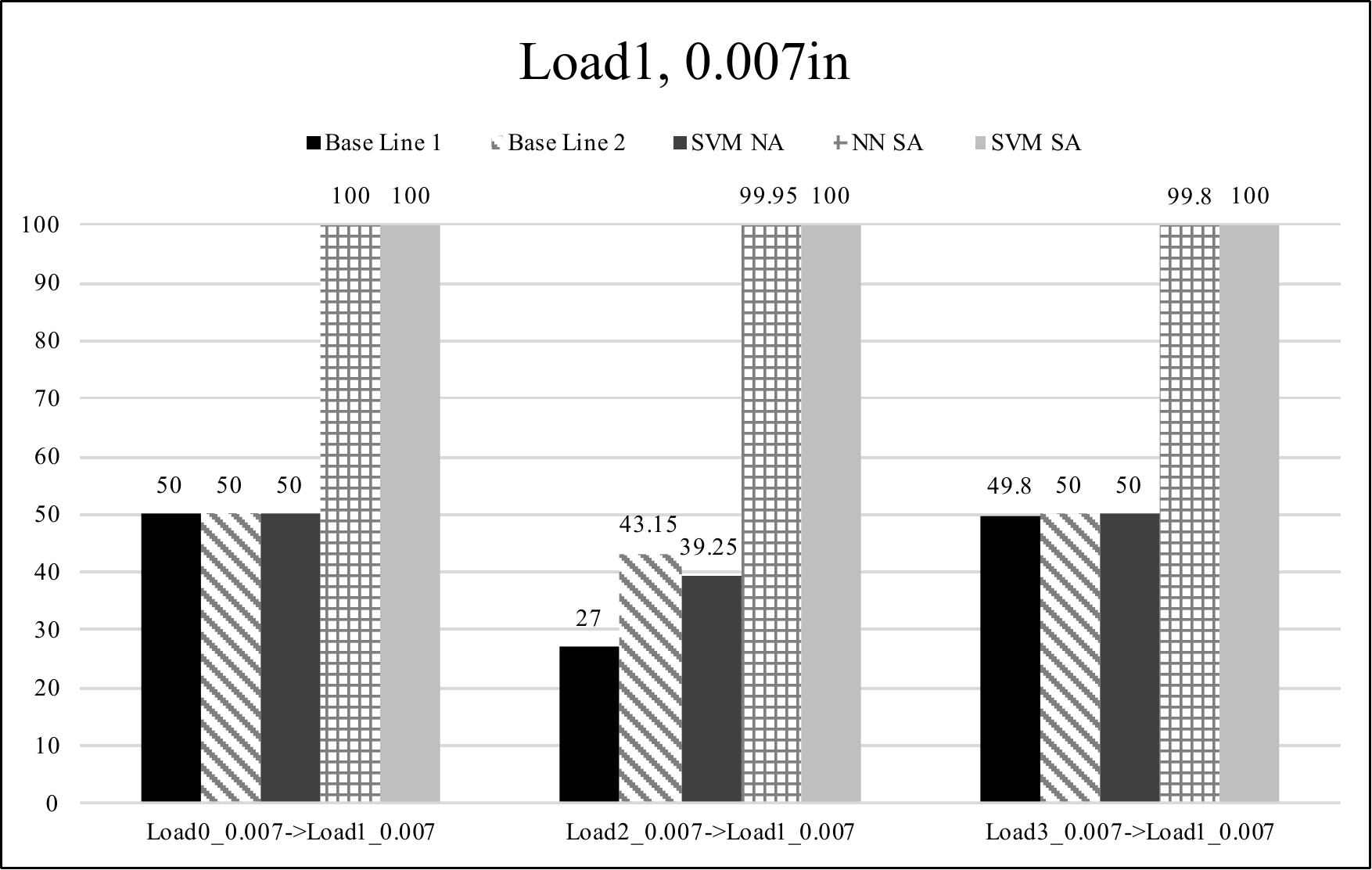}
  \label{fig_1_7_case}}
  \hfil
  \subfloat[]{\includegraphics[width=0.5\textwidth]{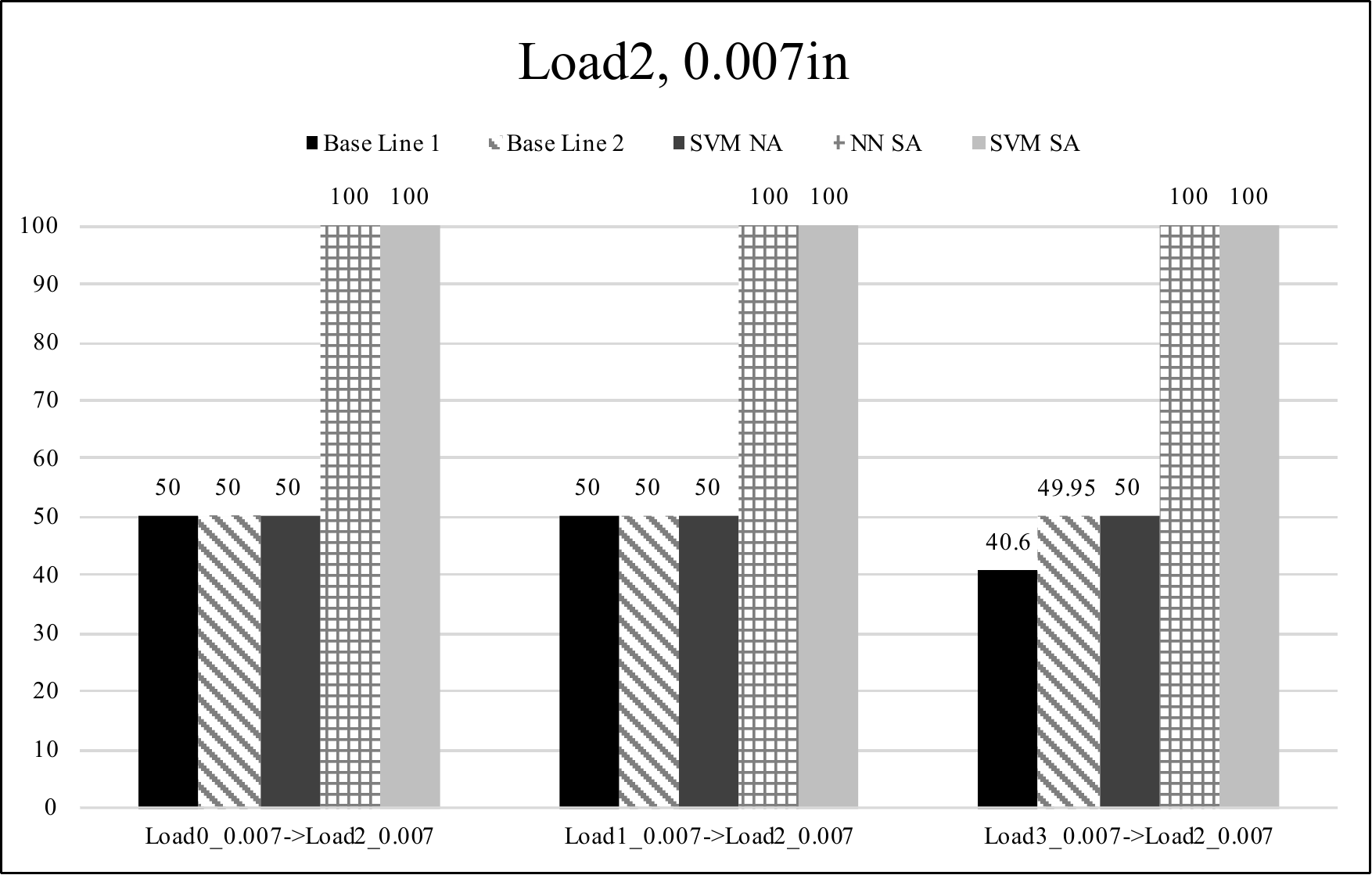}
  \label{fig_2_7_case}}
  \subfloat[]{\includegraphics[width=0.5\textwidth]{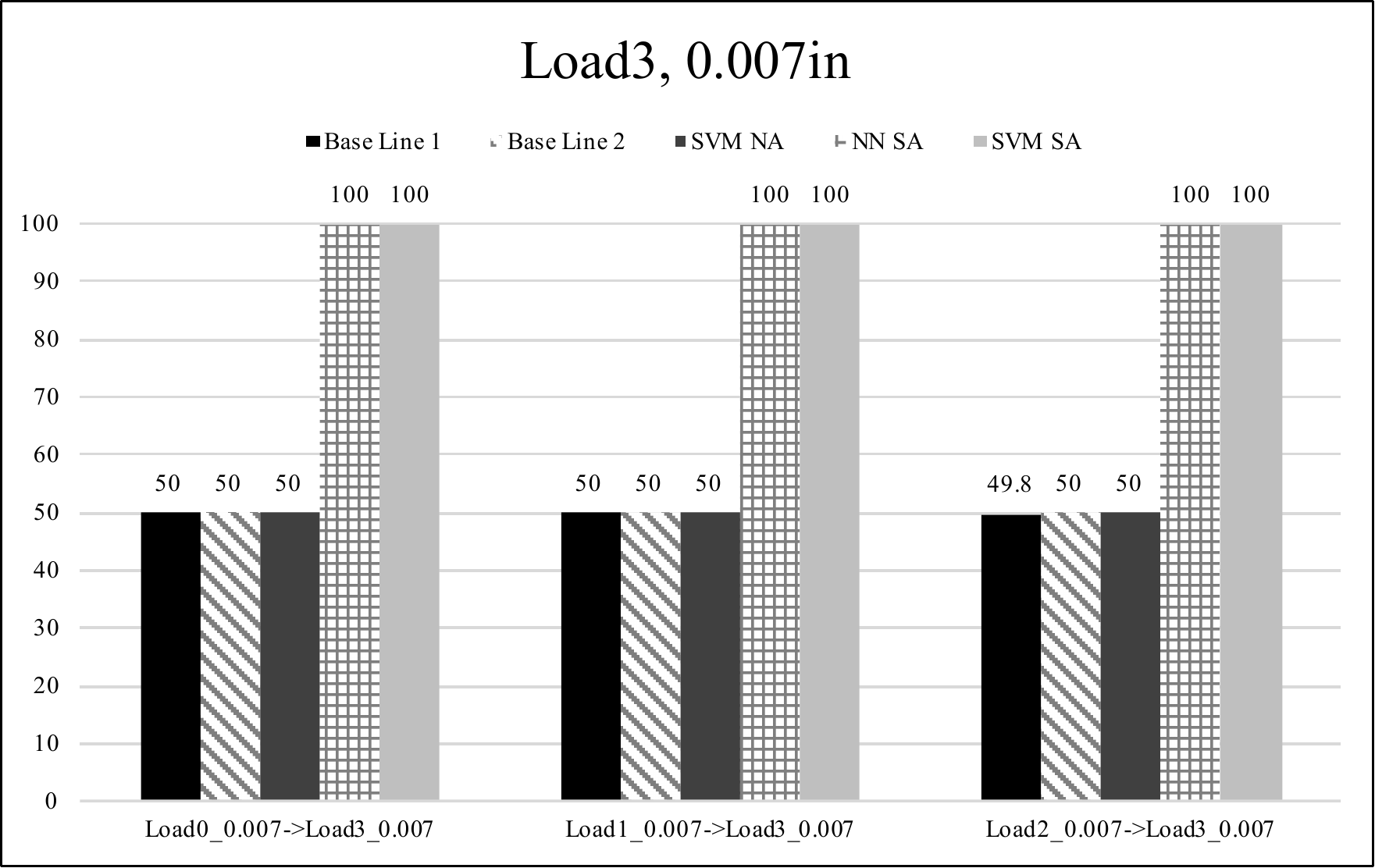}
  \label{fig_3_7_case}}
\caption{The results with fault size being 0.007in}
\label{Fig_case_0.007}
\end{figure*}

\begin{figure*}[!ht]
\centering
  \subfloat[]{\includegraphics[width=0.5\textwidth]{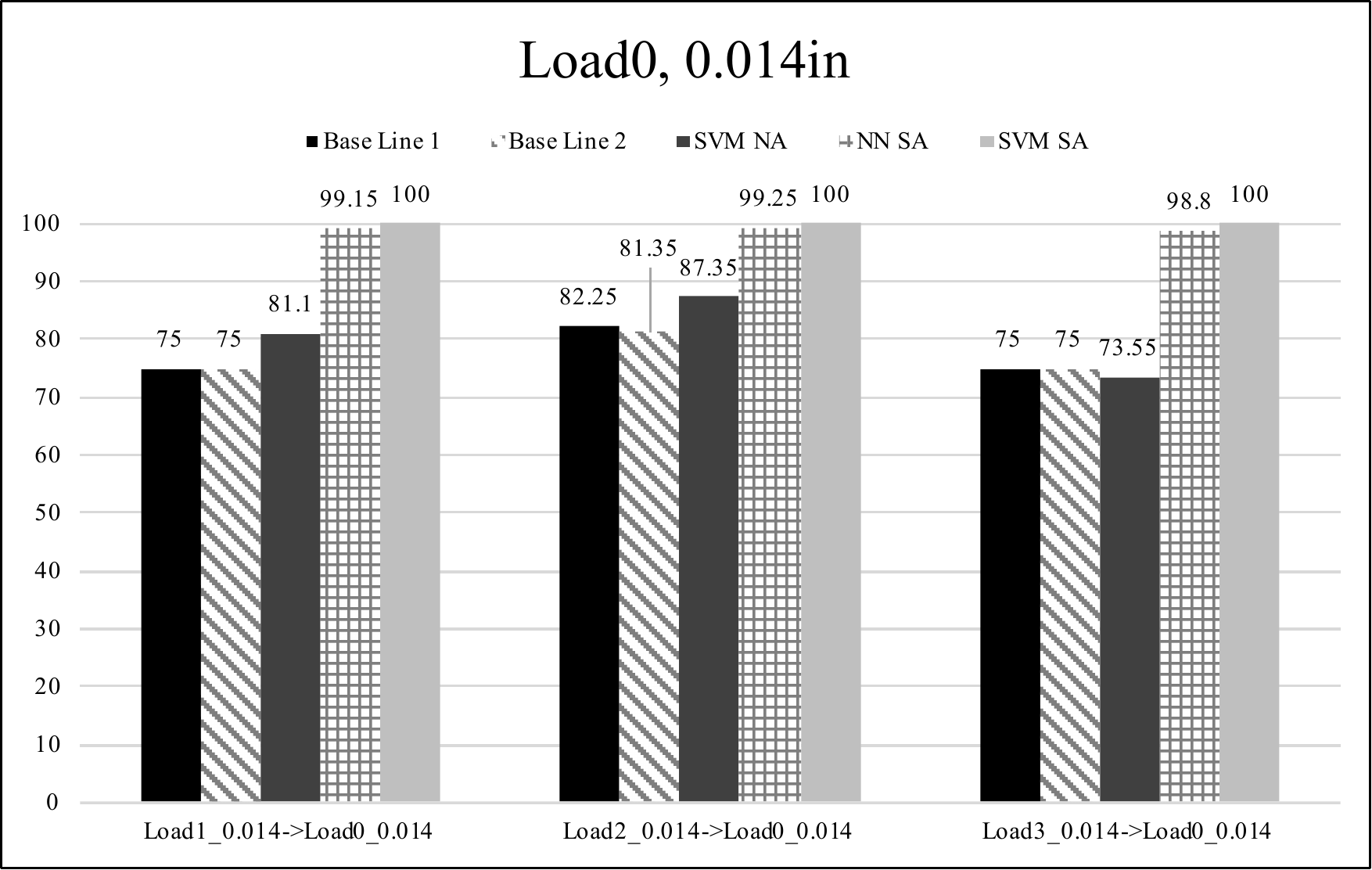}
  \label{fig_0_14_case}}
  \subfloat[]{\includegraphics[width=0.5\textwidth]{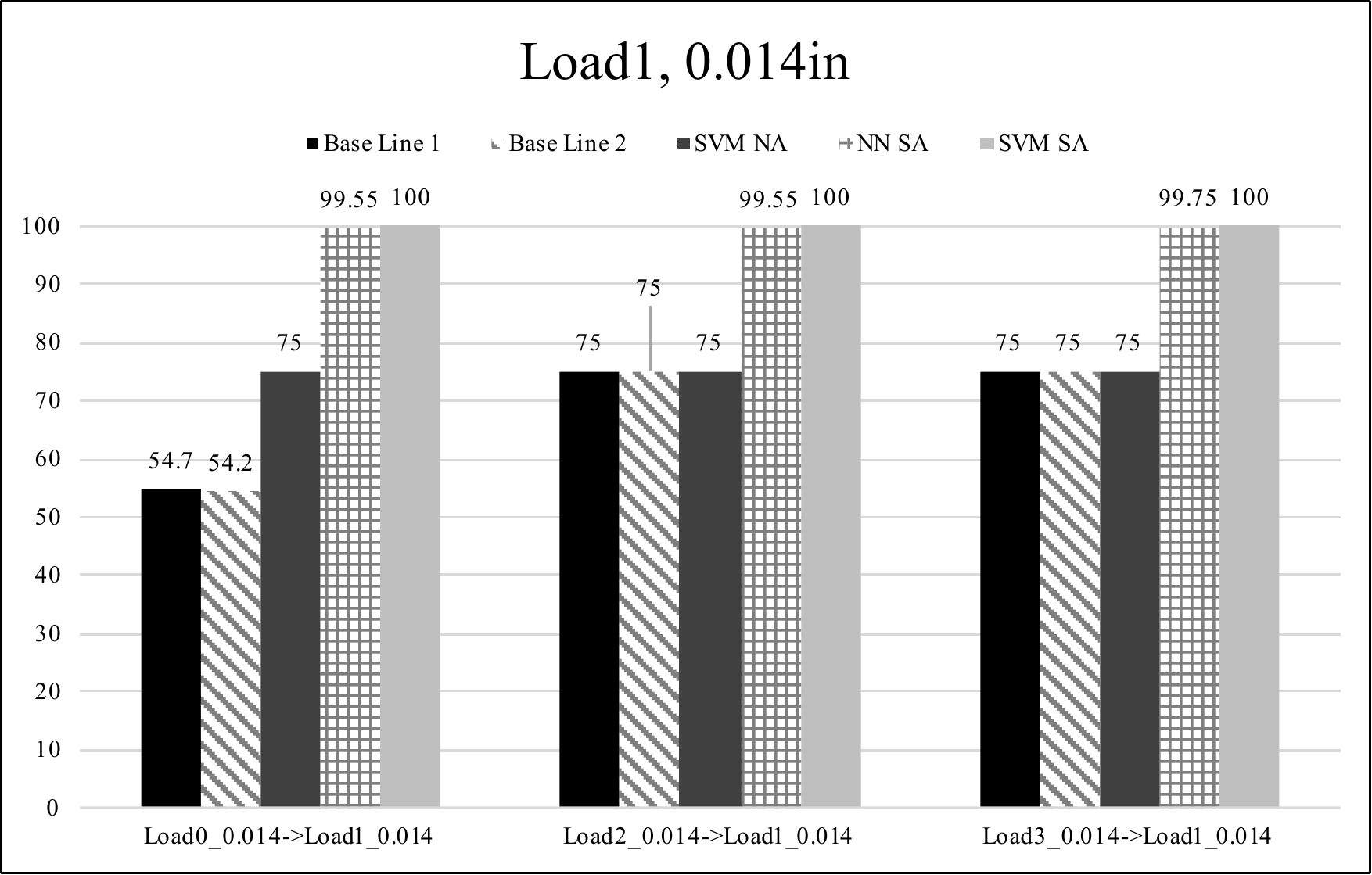}
  \label{fig_1_14_case}}
  \hfil
  \subfloat[]{\includegraphics[width=0.5\textwidth]{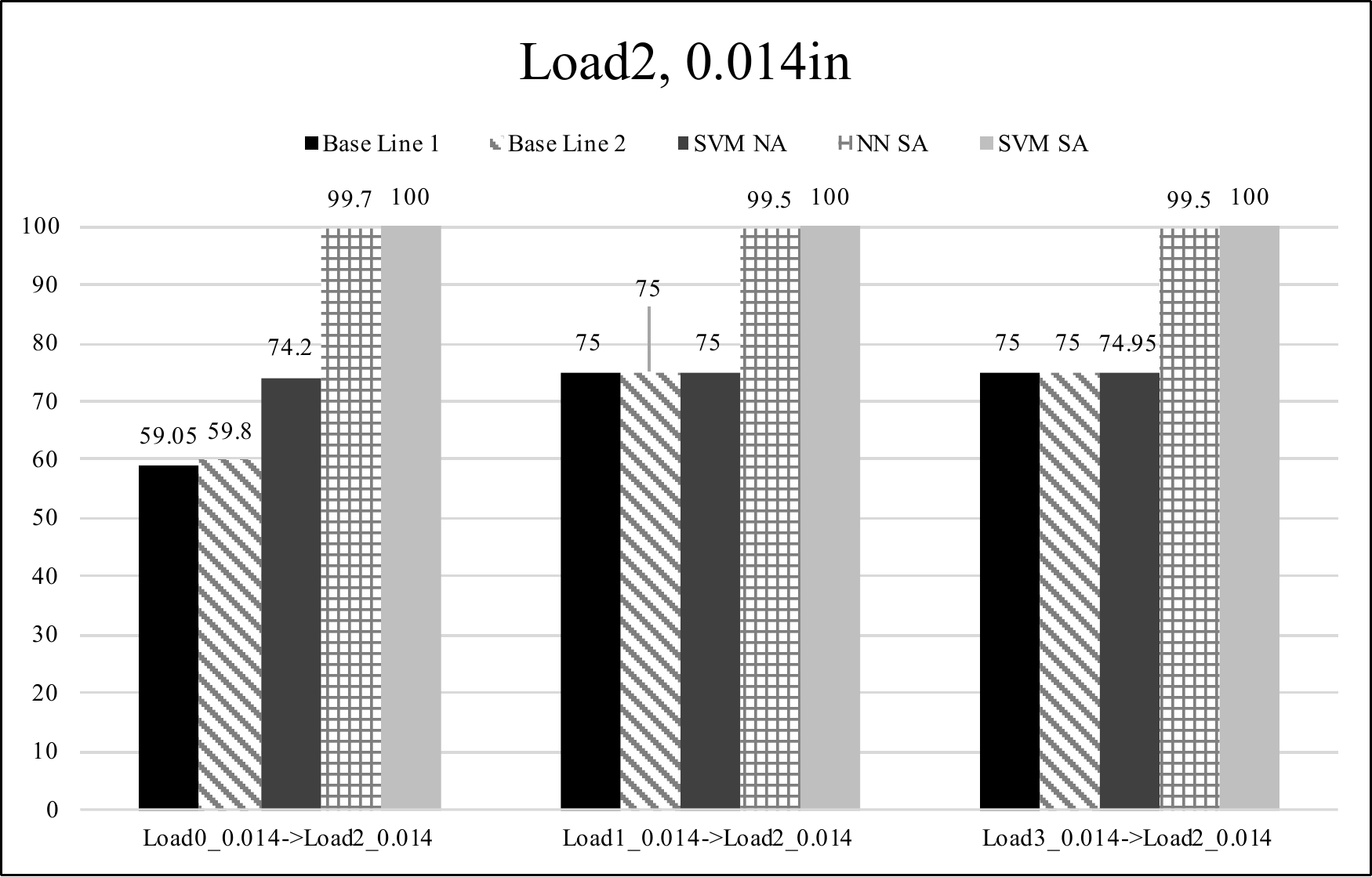}
  \label{fig_2_14_case}}
  \subfloat[]{\includegraphics[width=0.5\textwidth]{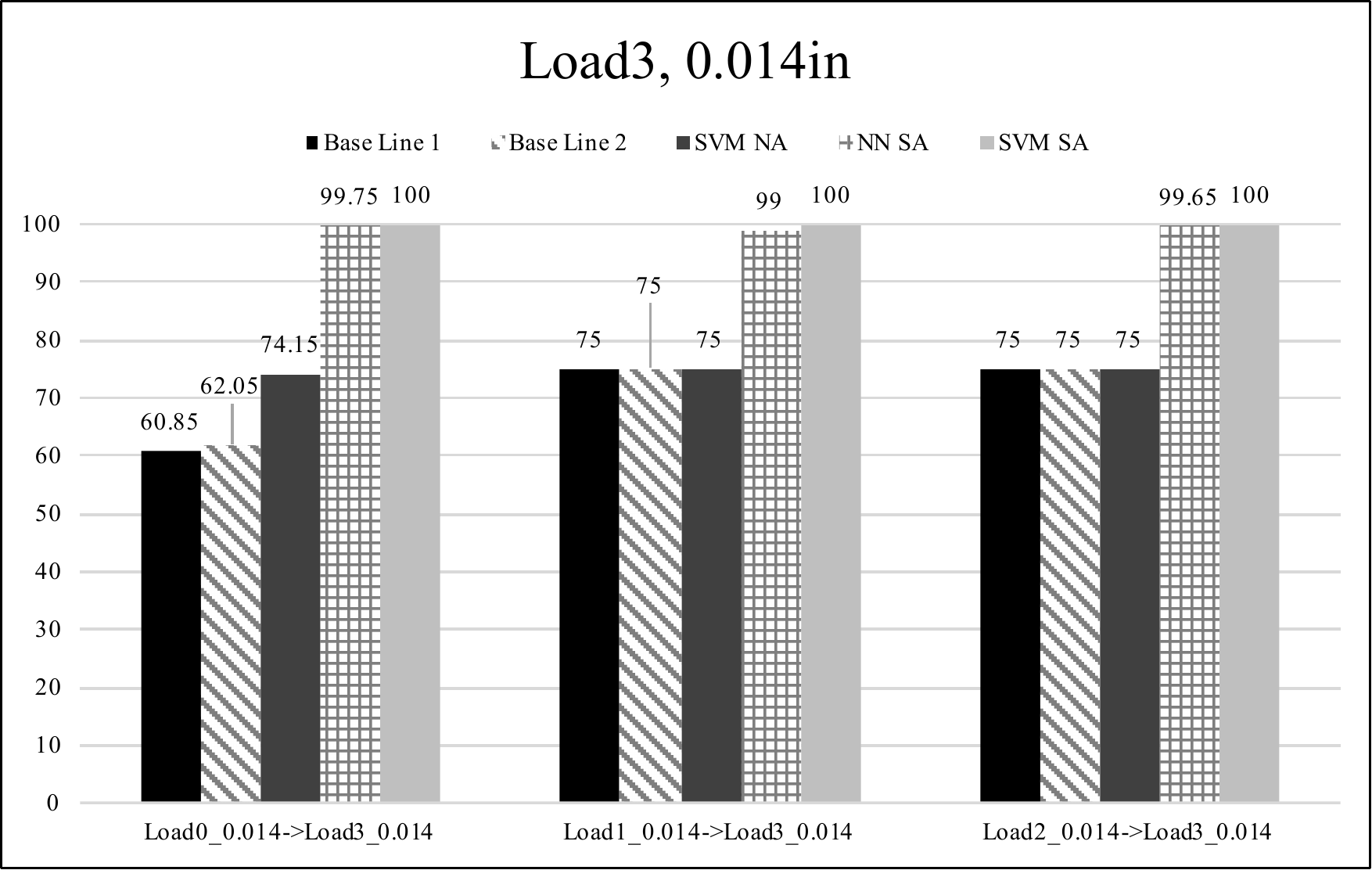}
  \label{fig_3_14_case}}
\caption{The results with fault size being 0.014in}
\label{Fig_case_0.014}
\end{figure*}

\begin{figure*}[!ht]
\centering
  \subfloat[]{\includegraphics[width=0.5\textwidth]{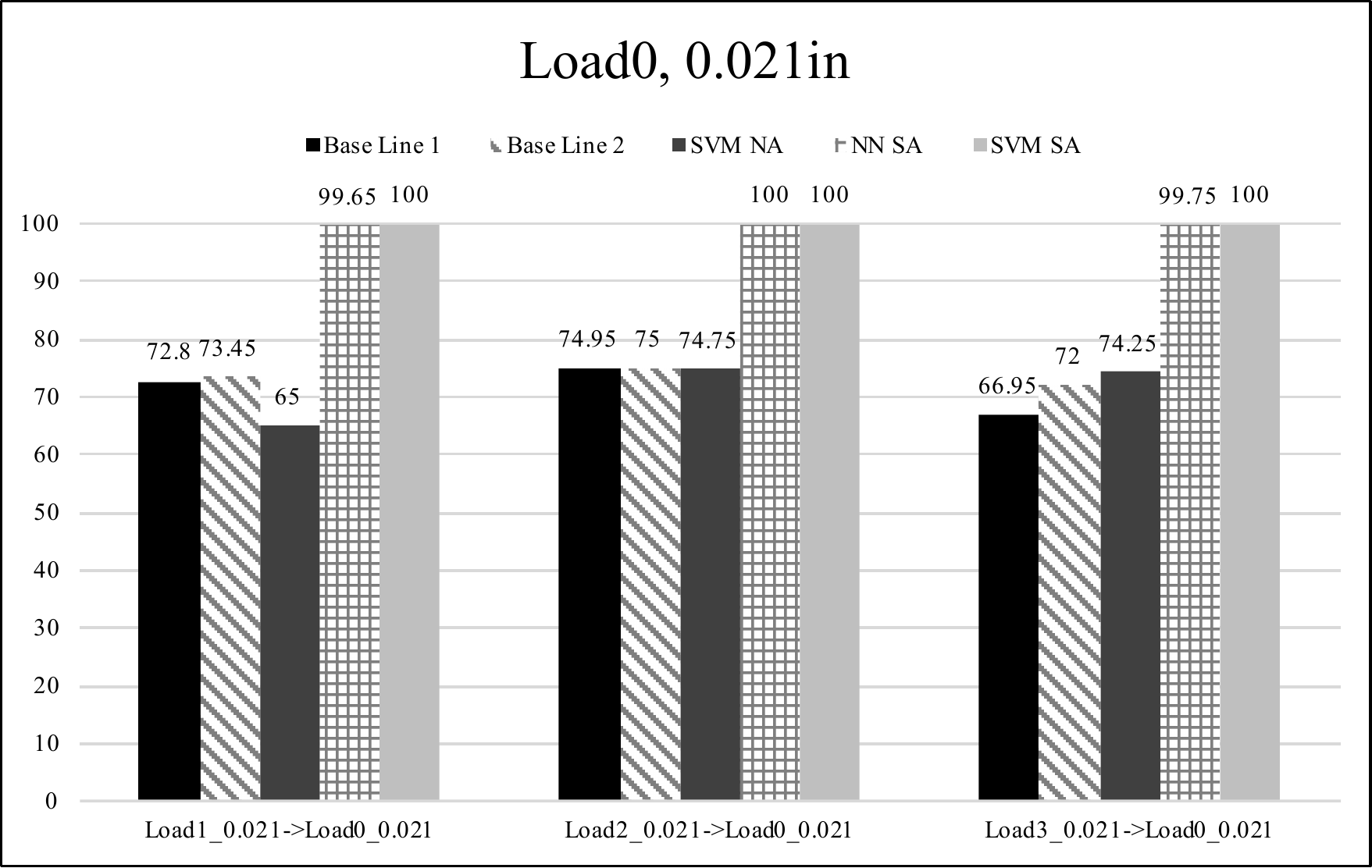}
  \label{fig_0_21_case}}
  \subfloat[]{\includegraphics[width=0.5\textwidth]{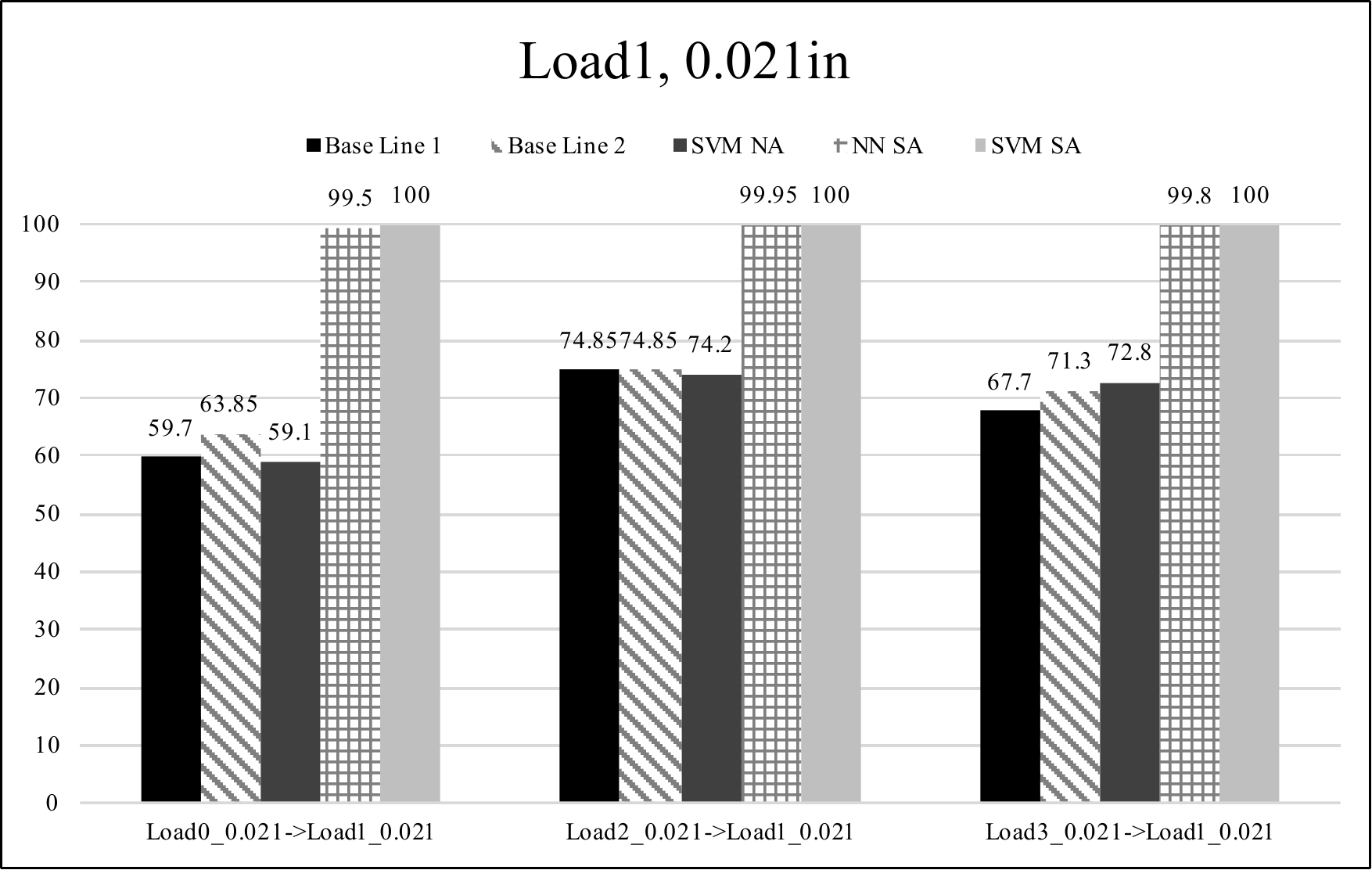}
  \label{fig_1_21_case}}
  \hfil
  \subfloat[]{\includegraphics[width=0.5\textwidth]{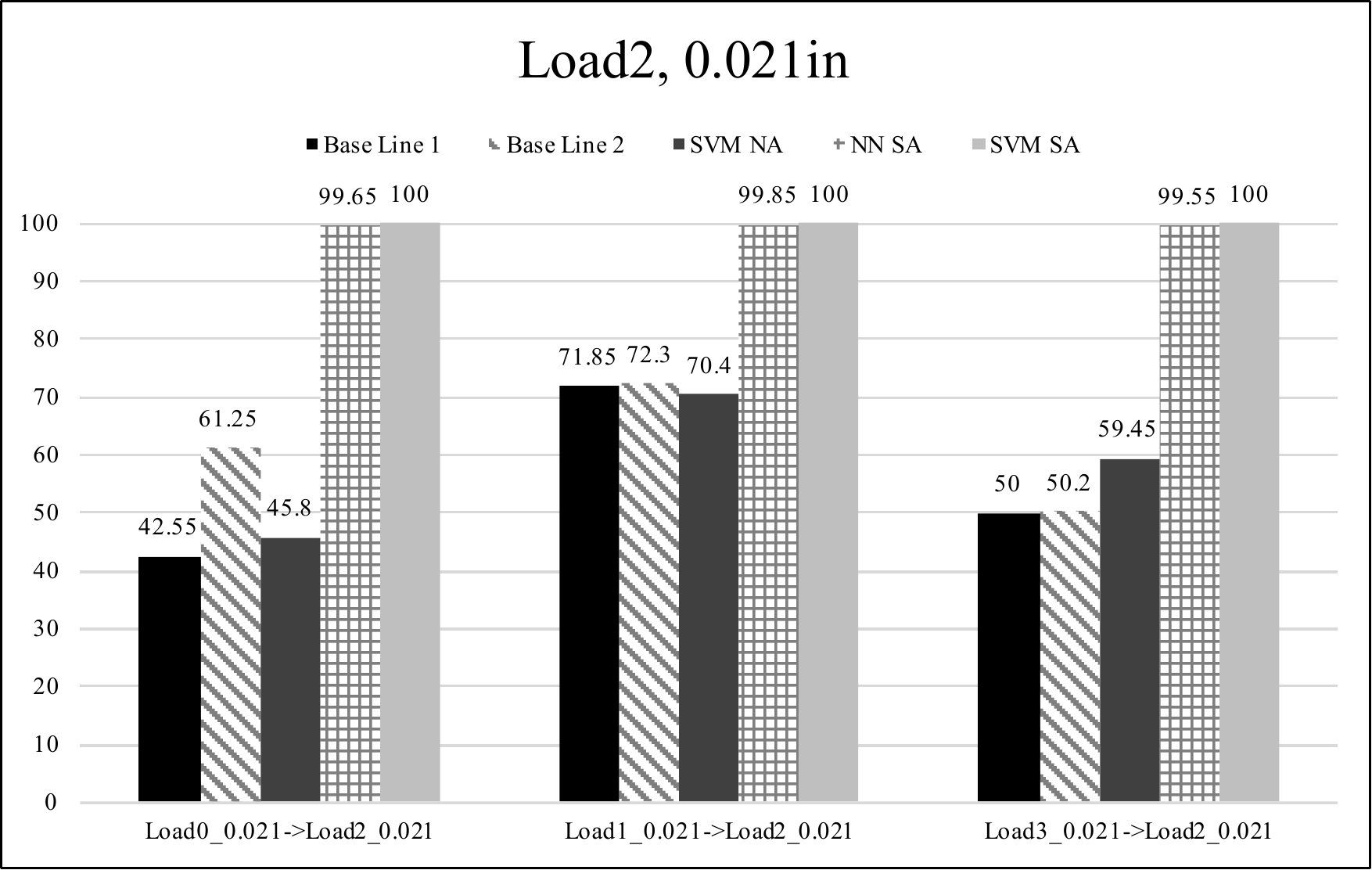}
  \label{fig_2_21_case}}
  \subfloat[]{\includegraphics[width=0.5\textwidth]{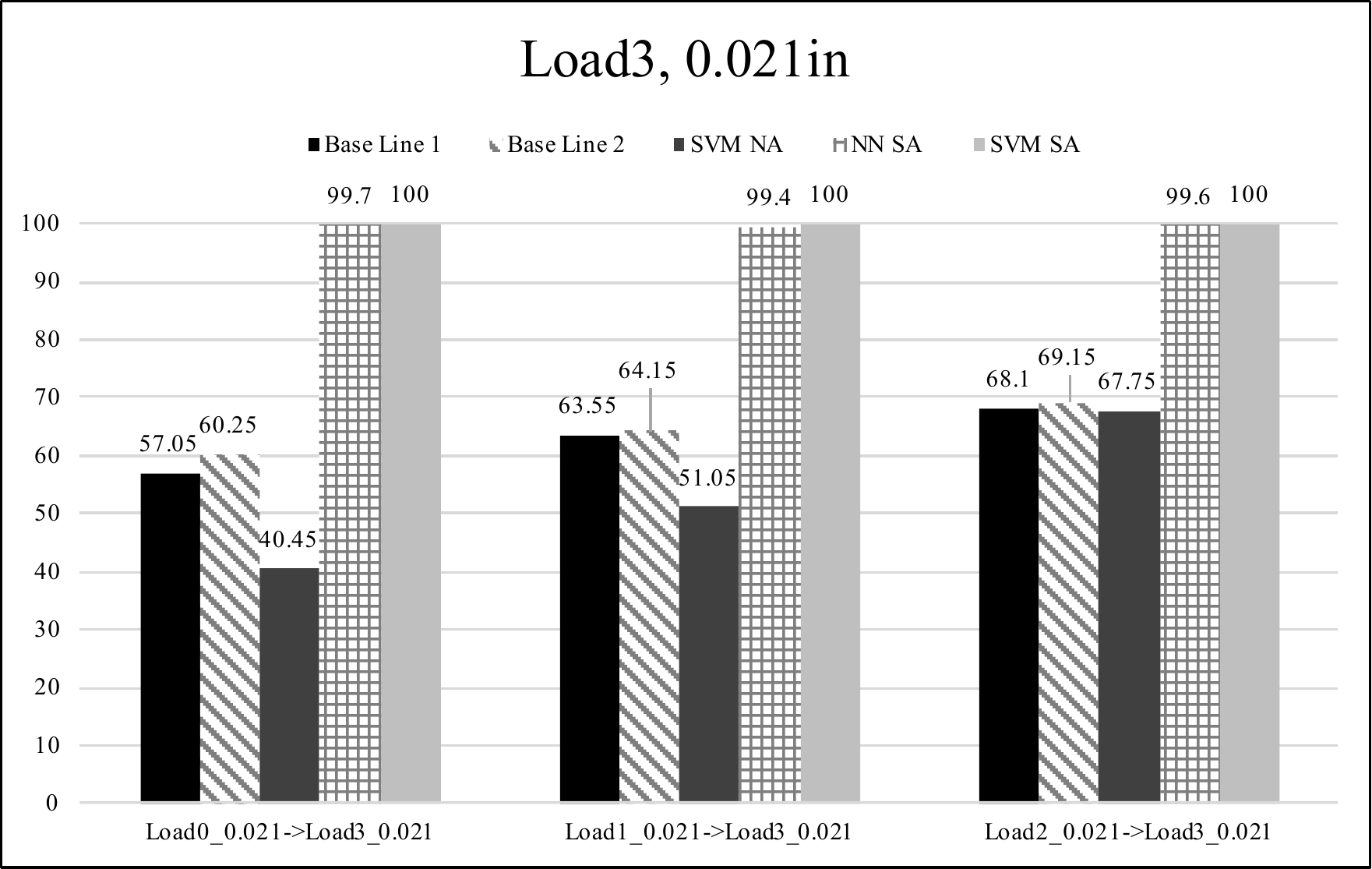}
  \label{fig_3_21_case}}
\caption{The results with fault size being 0.021in}
\label{Fig_case_0.021}
\end{figure*}

\begin{figure*}[!ht]
\centering
  \subfloat[]{\includegraphics[width=0.5\textwidth]{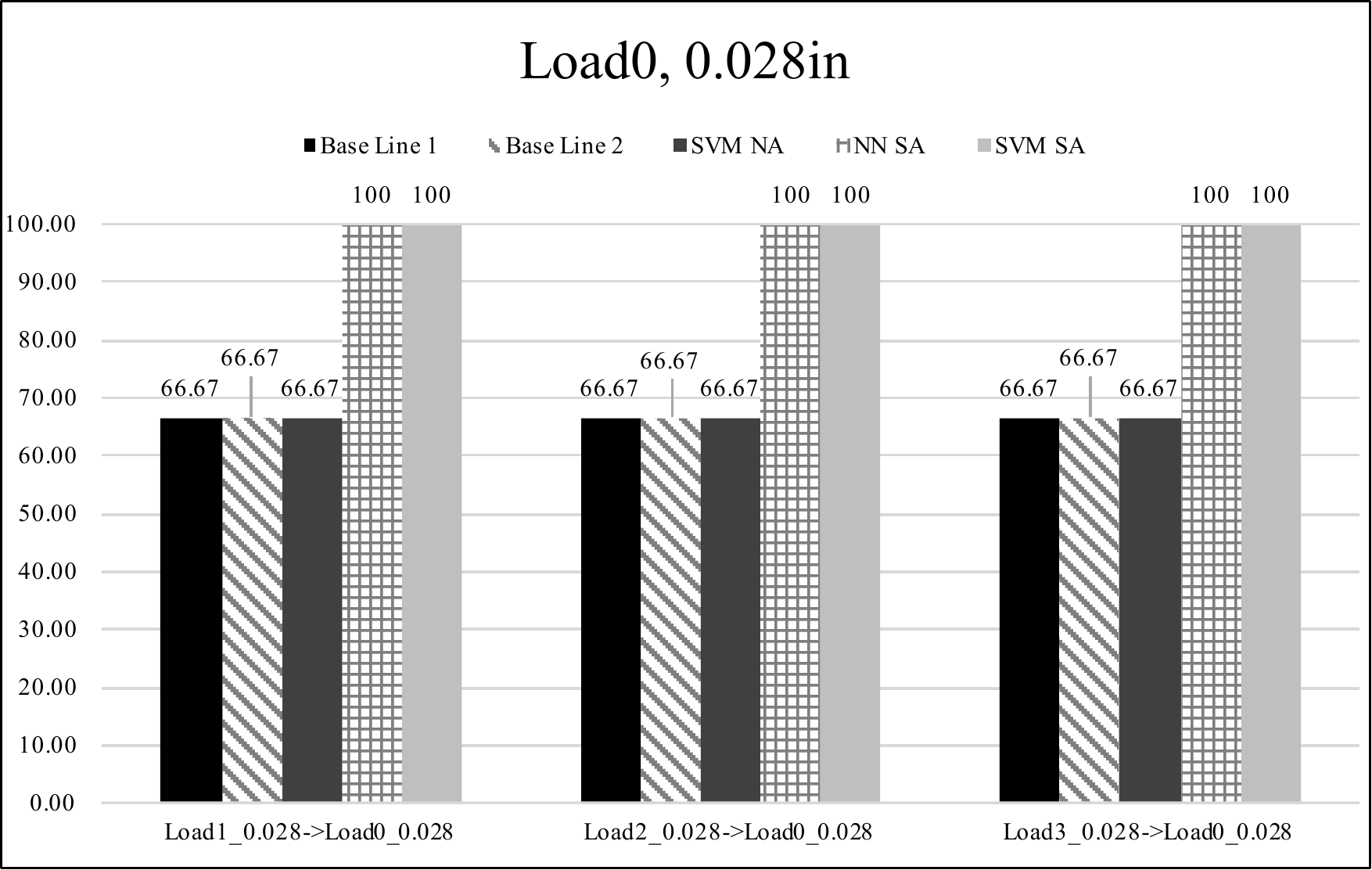}
  \label{fig_0_28_case}}
  \subfloat[]{\includegraphics[width=0.5\textwidth]{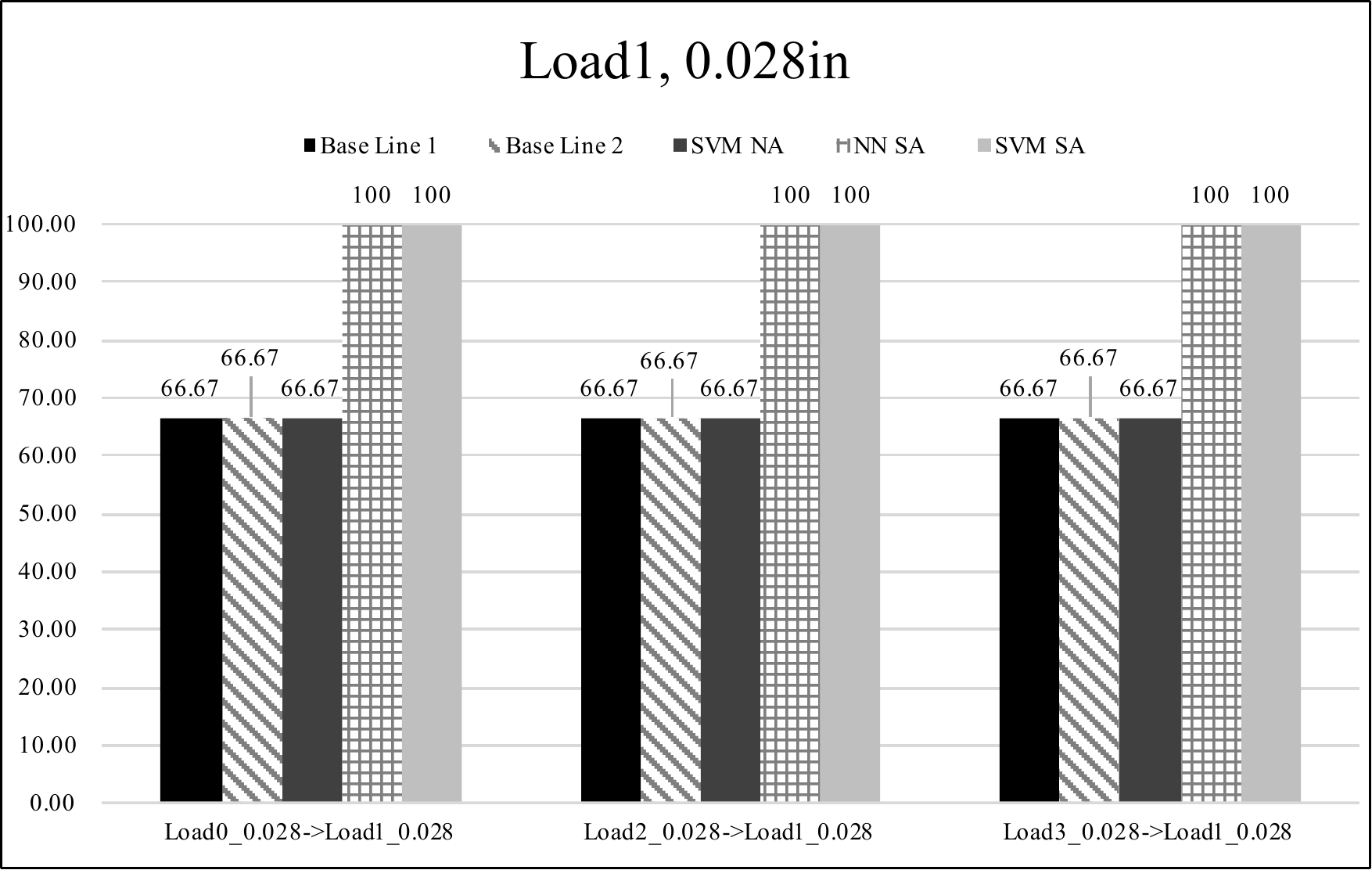}
  \label{fig_1_28_case}}
  \hfil
  \subfloat[]{\includegraphics[width=0.5\textwidth]{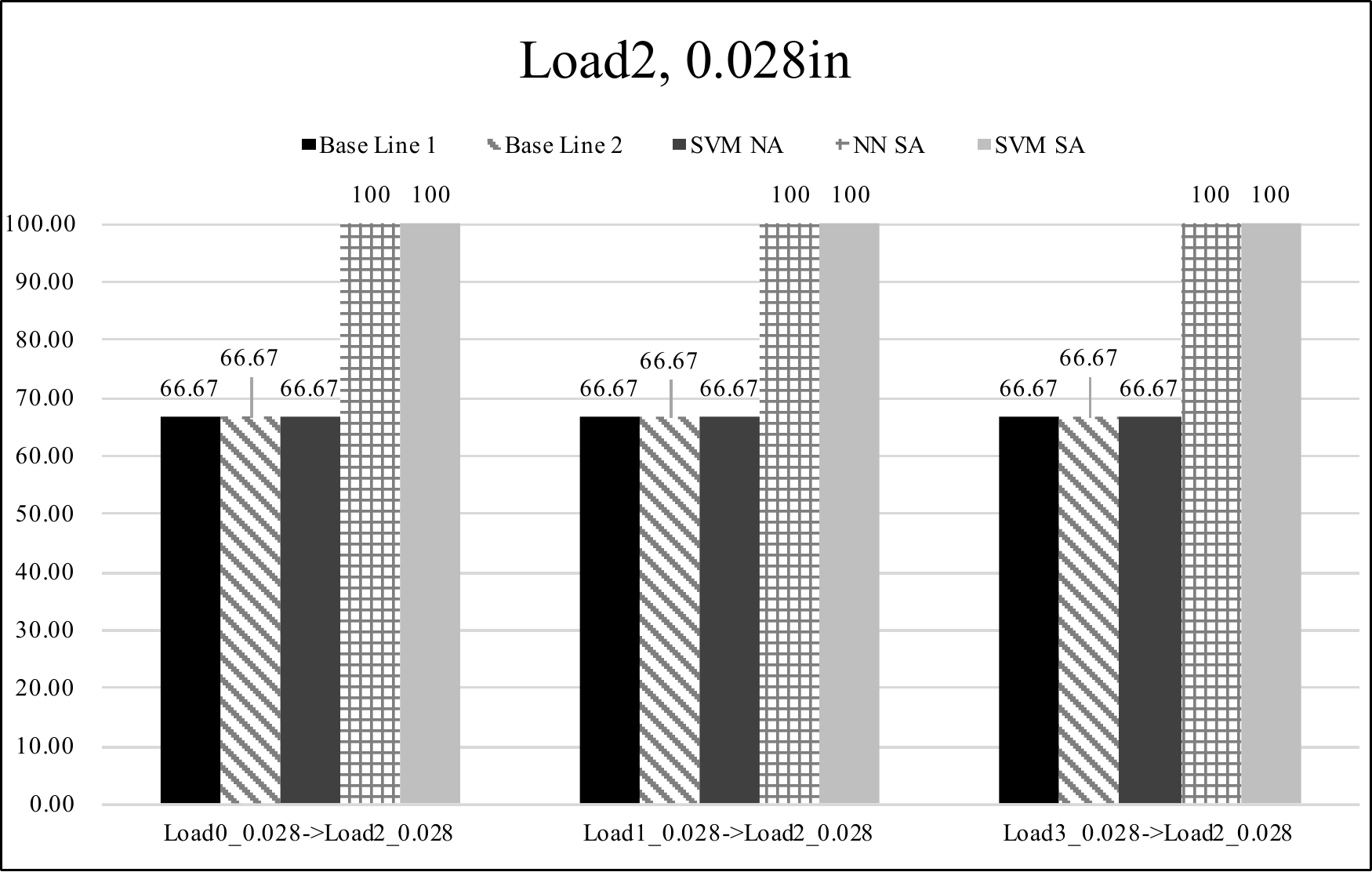}
  \label{fig_2_28_case}}
  \subfloat[]{\includegraphics[width=0.5\textwidth]{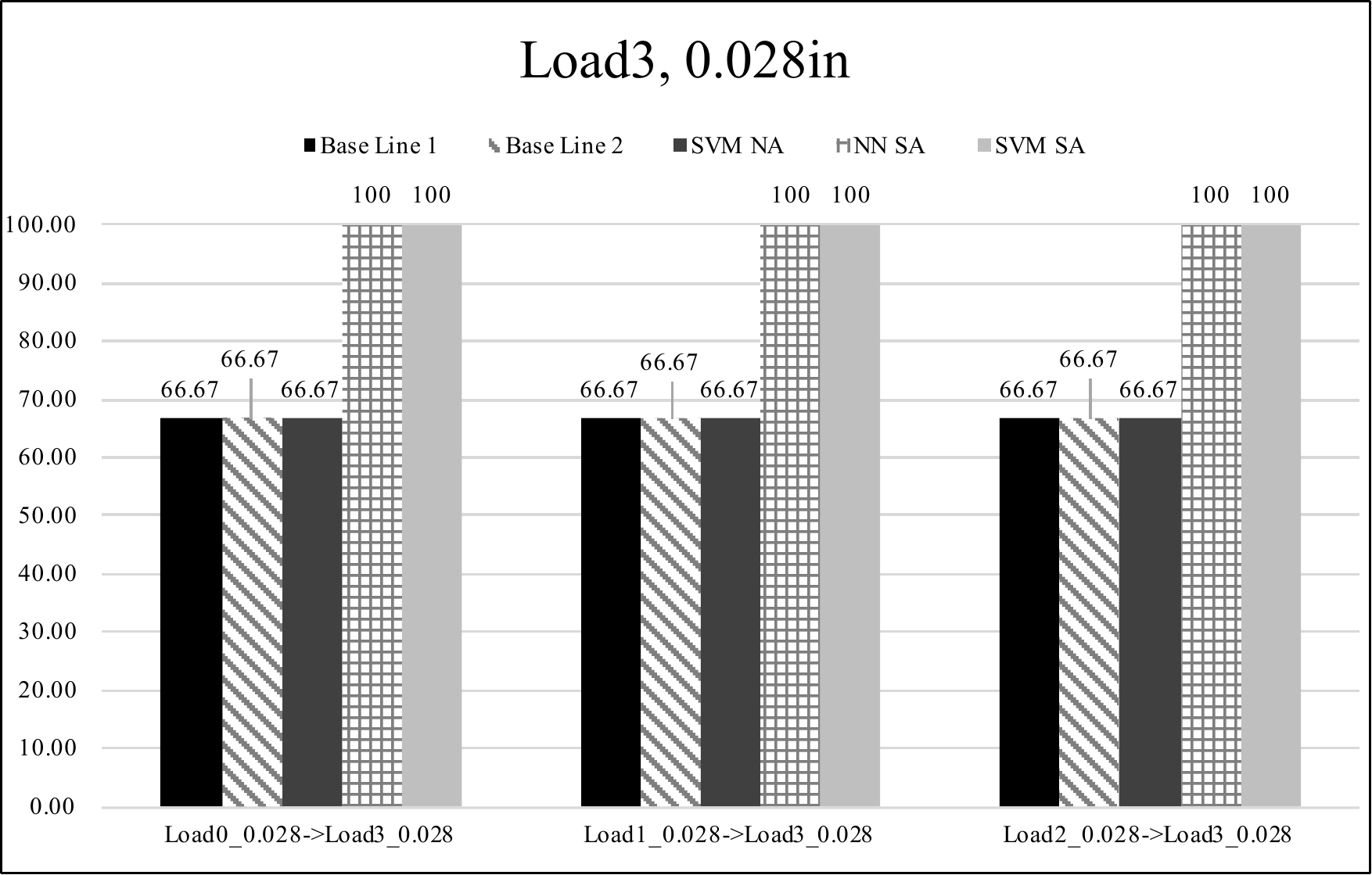}
  \label{fig_3_28_case}}
\caption{The results with fault size being 0.028in}
\label{Fig_case2_0.028}
\end{figure*}

The diagnosis results of five methods with fault size being 0.007in, 0.014in, 0.021in, and 0.028in are illustrated in Figure \ref{Fig_case_0.007}, Figure \ref{Fig_case_0.014}, Figure \ref{Fig_case_0.021} and Figure \ref{Fig_case2_0.028} respectively. Each figure consists of four parts, which have the same test domain, and each figure has the same fault size. For each set of bars, the left of the symbol $"->"$ represents the training domain and the right represents testing domain. For example, in Figure \ref{Fig_case_0.007}, the test domains are ordered clockwise from the top left: Load0\_0.007, Load1\_0.007, Load2\_0.007, and Load3\_0.007. In Figure \ref{fig_0_7_case}, one part of Figure \ref{Fig_case_0.007},  the testing domain is Load0\_0.007, and the training domains are Load1\_0.007, Load2\_0.007, Load3\_0.007 respectively. From the poor performance of methods Baseline 1, Baseline 2 and SVM NA, in these four figures, we can find that the distributions of training data and testing data are different. As the four figures show, comparing the five methods, we can find that the accuracy of proposed method, SVM SA, are all $100\%$. That's really exciting. It's also obvious that although the proposed method NN SA is little inferior to SVM SA, the accuracies are all close to $100\%$ and the performance of NN NA is obviously better than other three methods. So the proposed two methods are more domain invariant than the traditional methods.

\begin{figure*}[!ht]
\centering
  \subfloat[No Adaption, $H{\Delta}H$=2.0, accuracy=$47.05\%$]{\includegraphics[width=0.33\textwidth]{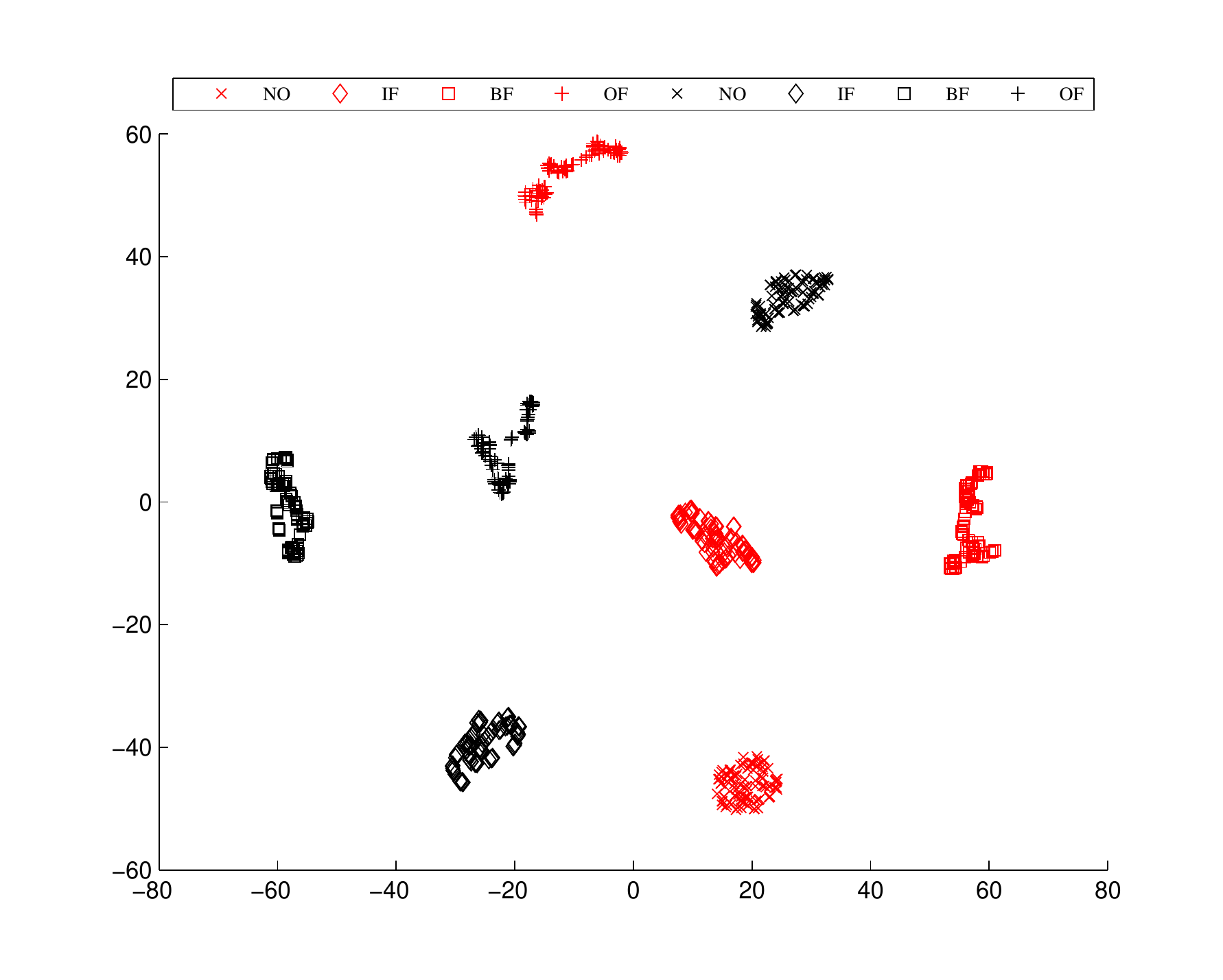}
  \label{tSNE_7_2_0_NA}}
  \subfloat[PCA, $H{\Delta}H$=2.0, accuracy=$50\%$]{\includegraphics[width=0.33\textwidth]{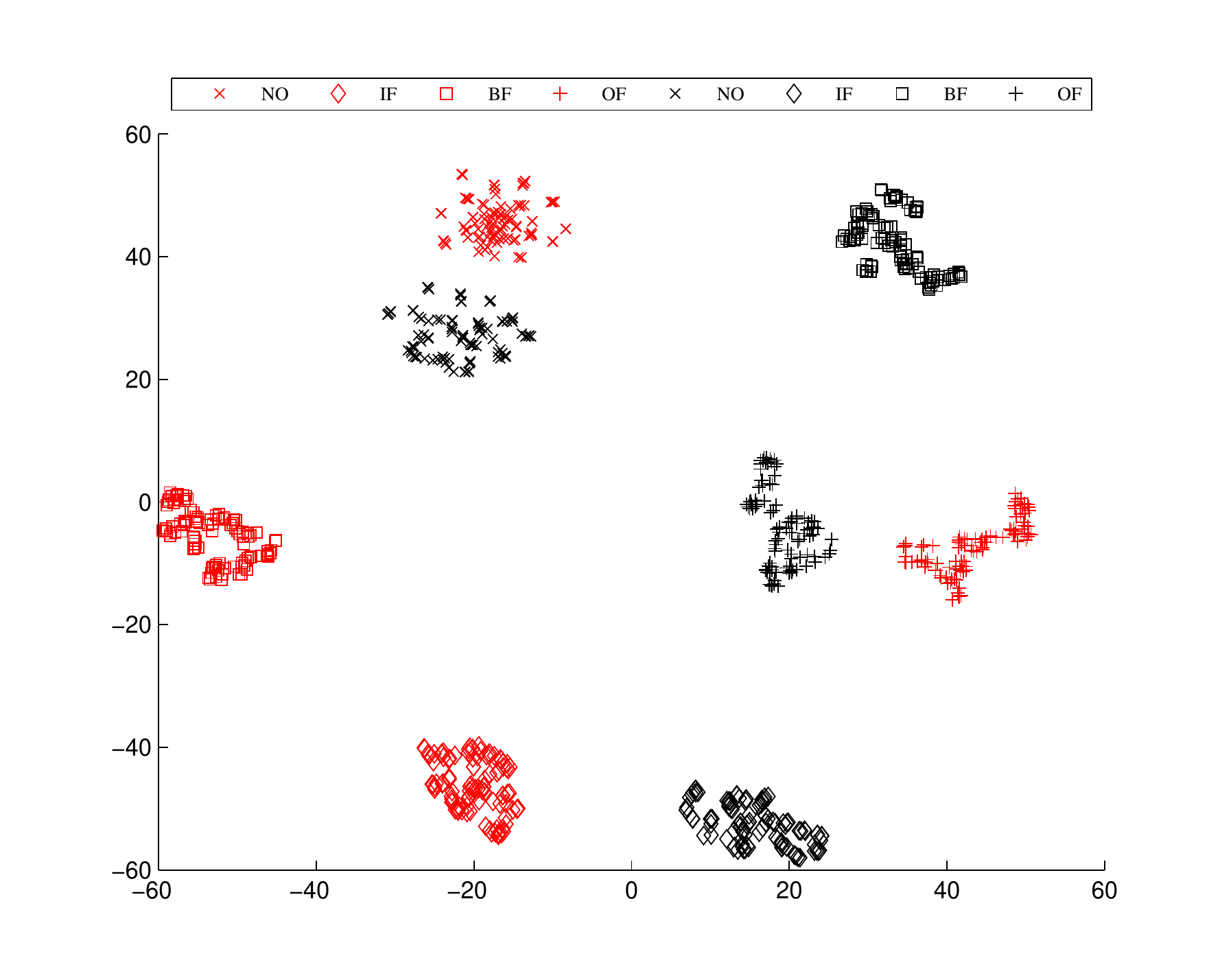}
  \label{tSNE_7_2_0_PCA}}
  \subfloat[Unsupervised DA with SA, $H{\Delta}H$=0.12, accuracy=$99.85\%$]{\includegraphics[width=0.33\textwidth]{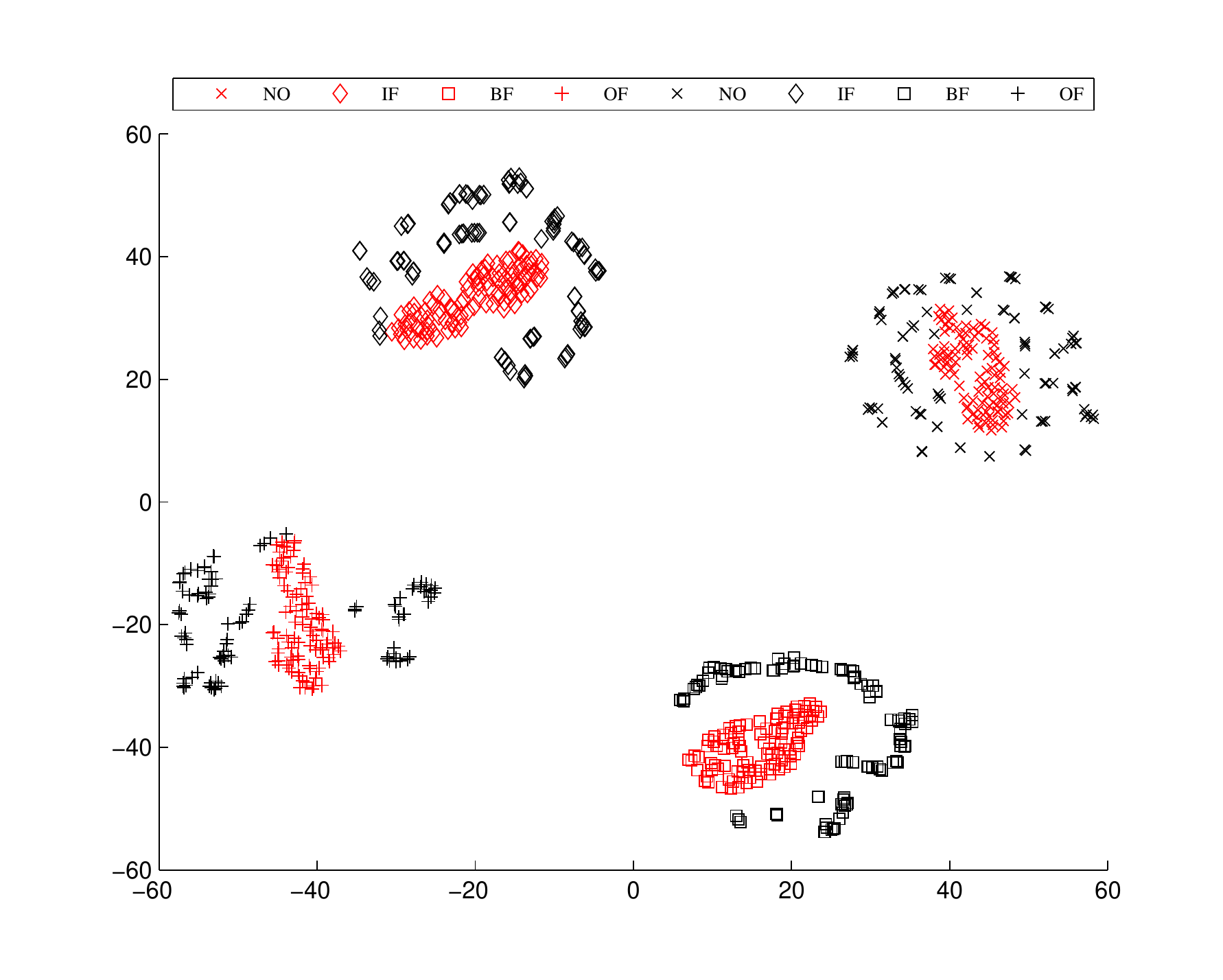}
  \label{tSNE_7_2_0_SA}}
\caption{Data visualization using t-SNE \cite{Laurens2008Visualizing} over a typical domain adaption task from Load2\_0.007(red) to Load0\_0.007(black) in dataset A. $H{\Delta}H$ and recognition accuracy are indicated in each sub-figure title.}
\label{Fig_distribution_Load0_Load2}
\end{figure*}

In order to investigate why the proposed method can improve the classification performance, t-SNE \cite{Laurens2008Visualizing} is utilized to visualize the feature representation distribution of Load0\_0.007 and Load2\_0.007 with fault size being 0.007in, after the process of three methods, no adaptation, PCA and unsupervised DA with SA, as the Figure \ref{Fig_distribution_Load0_Load2} shows. It's just a random example. It is clear that the domain discrepancy between two datasets after unsupervised DA with SA is obviously smaller than after the other two methods. The lower the $H{\Delta}H$ is, the better two distributions align and the higher the accuracy is. So, in the proposed methods the classifier trained on Load0 with fault size being 0,007in can be used to classify the data in Load2 with fault size being 0.007in. That's why the high accuracy occurs.

\begin{figure*}[!ht]
\centering
  \subfloat[]{\includegraphics[width=0.5\textwidth]{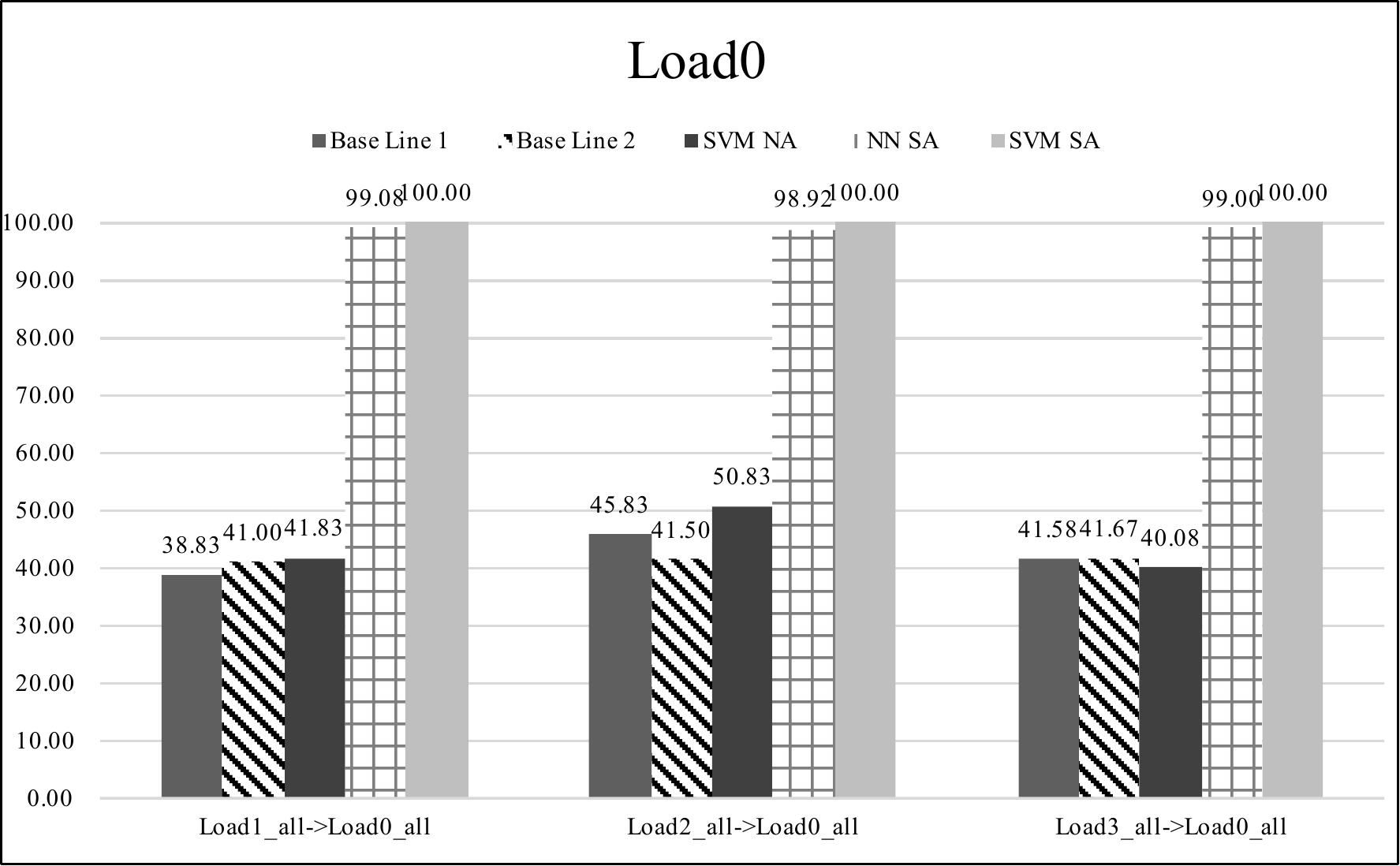}
  \label{fig_0_all_case}}
  \subfloat[]{\includegraphics[width=0.5\textwidth]{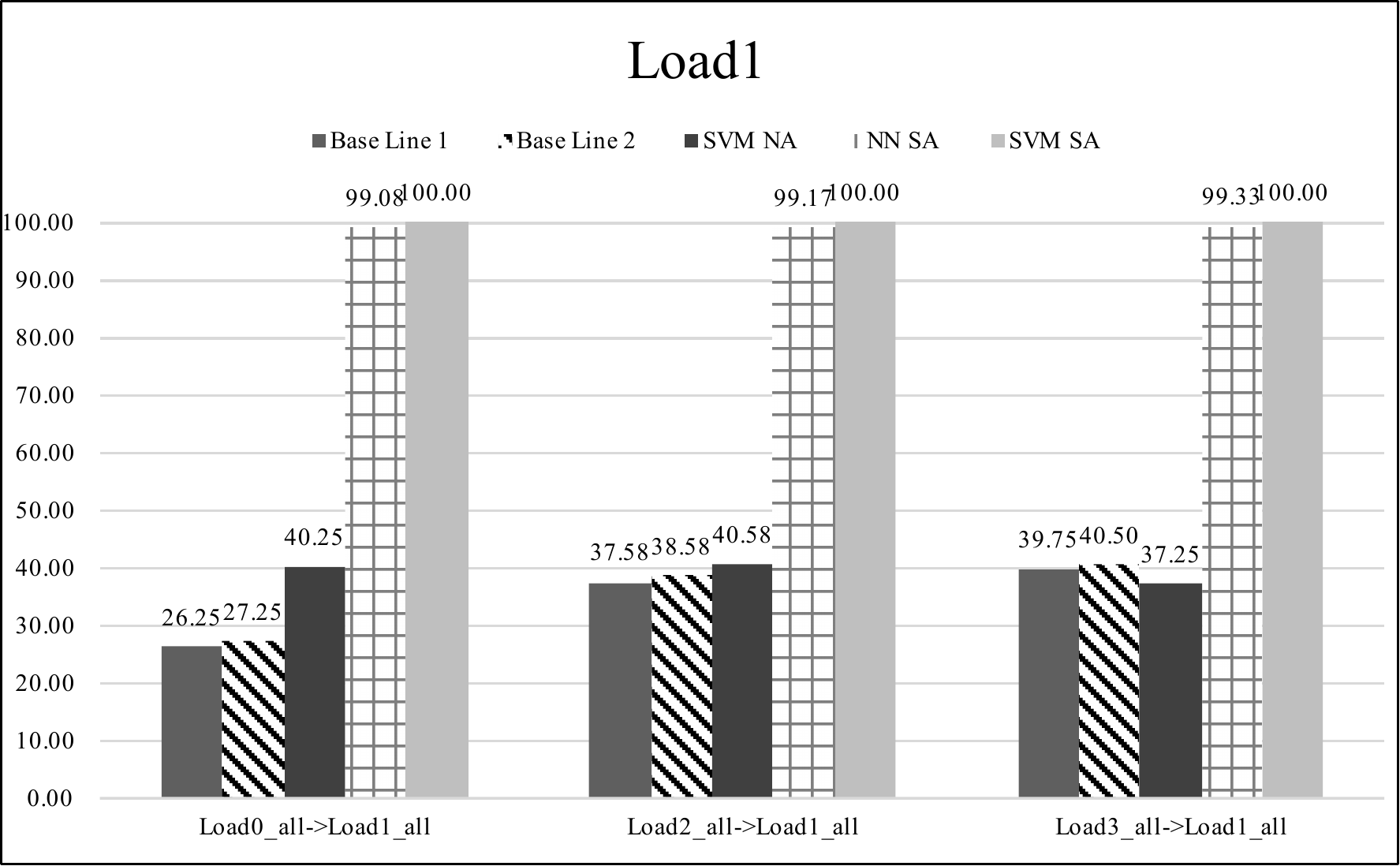}
  \label{fig_1_all_case}}
  \hfil
  \subfloat[]{\includegraphics[width=0.5\textwidth]{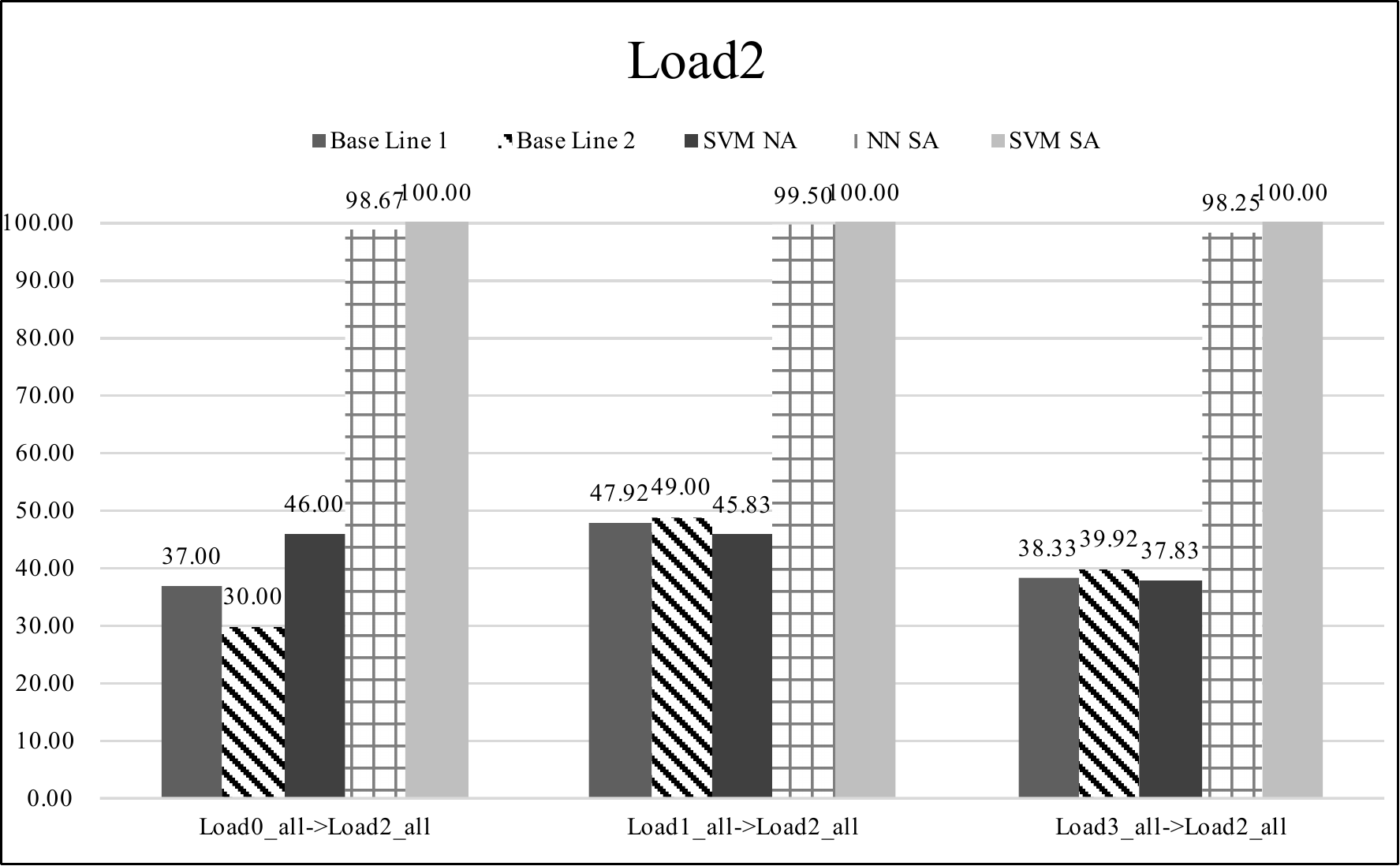}
  \label{fig_2_all_case}}
  \subfloat[]{\includegraphics[width=0.5\textwidth]{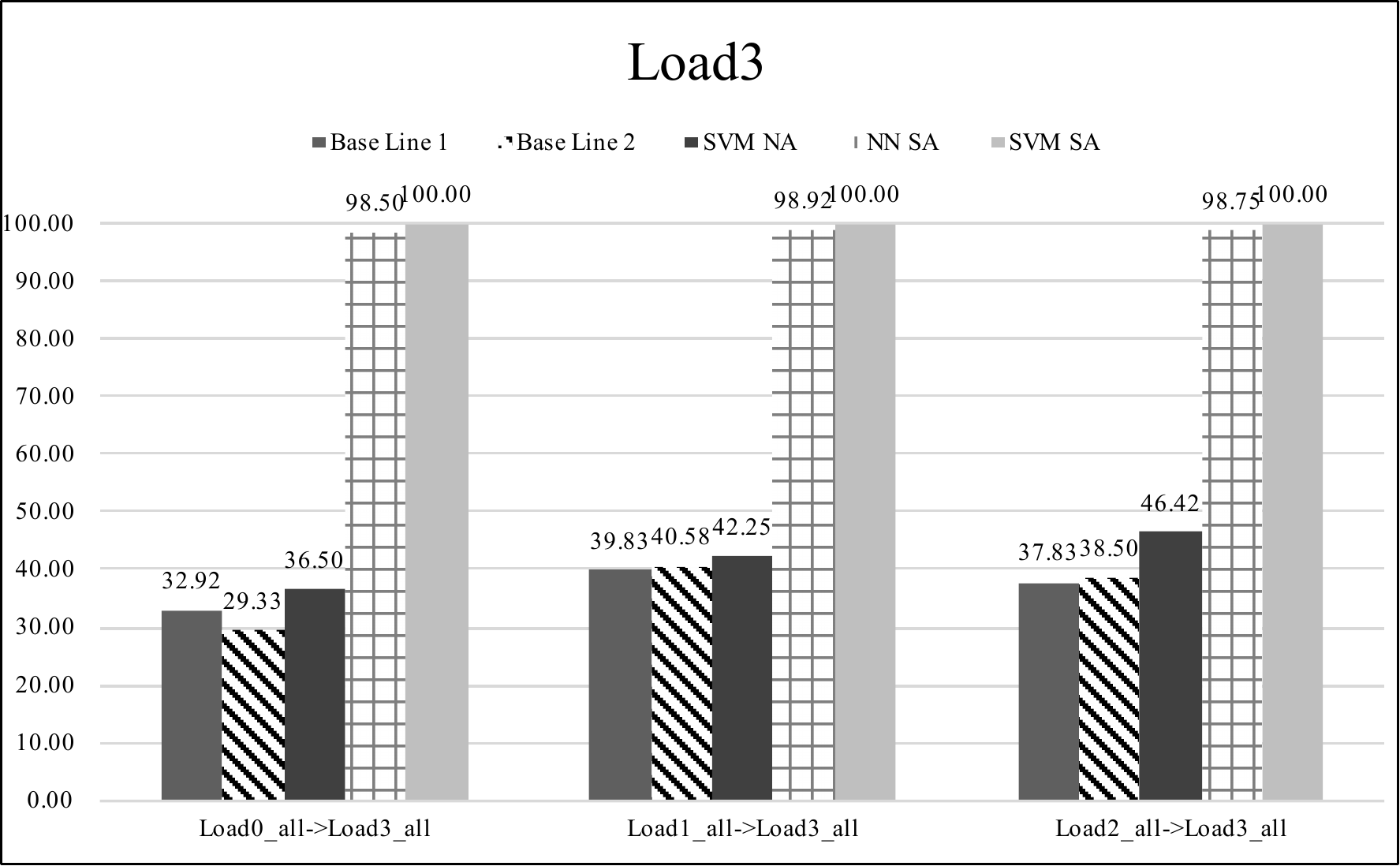}
  \label{fig_3_all_case}}
\caption{The results with fault size being 0.007in, 0.014in, 0.021in and 0.028in}
\label{Fig_case_all_load}
\end{figure*}

The database A also can be divided into four datasets, namely, Load0\_all, Load1\_all, Load2\_all and Load3\_all. Each dataset consists of 12 health condition data (There are IF, OF and BF with the fault size being 0.007in, 0.014in and 0.021in. There are IF, BF with fault size being 0.028in. Normal condition is also considered) under one certain working condition. Each sample is composed by the same way with the above experiment. So, there are other 12 domain adaptation diagnosis problems under varying working condition in Case Western Reserve benchmark data. The results are shown in Figure \ref{Fig_case_all_load}. For each set of bars, the test domains are same and the left and right of the symbol $"->"$ represents the training domain and testing domain respectively. These four figures show that the performance of the proposed method, SVM SA is wonderful and the accuracies are all $100\%$, which is far superior to the methods, Baseline 1, Baseline 2, and SVM NA. The proposed method, NN SA is also very good, although it is little inferior to SVM SA. We can conclude that the proposed methods can effectively distinguish not only bearing faults categories but also fault severities. It's also important to note that the proposed methods can be applied into the fault size being 0.028in which is little considered in other literatures. We should also note that during the train of classifier, the testing domain is totally unlabeled, which is very meaningful, because labeling the data is very hard in practice.

\subsection{Case 2: Fault diagnosis based on Database B}
\subsubsection{Experimental setup and database preparation}
\begin{figure*}[!ht]
  \centering
  \includegraphics[width=0.6\textwidth]{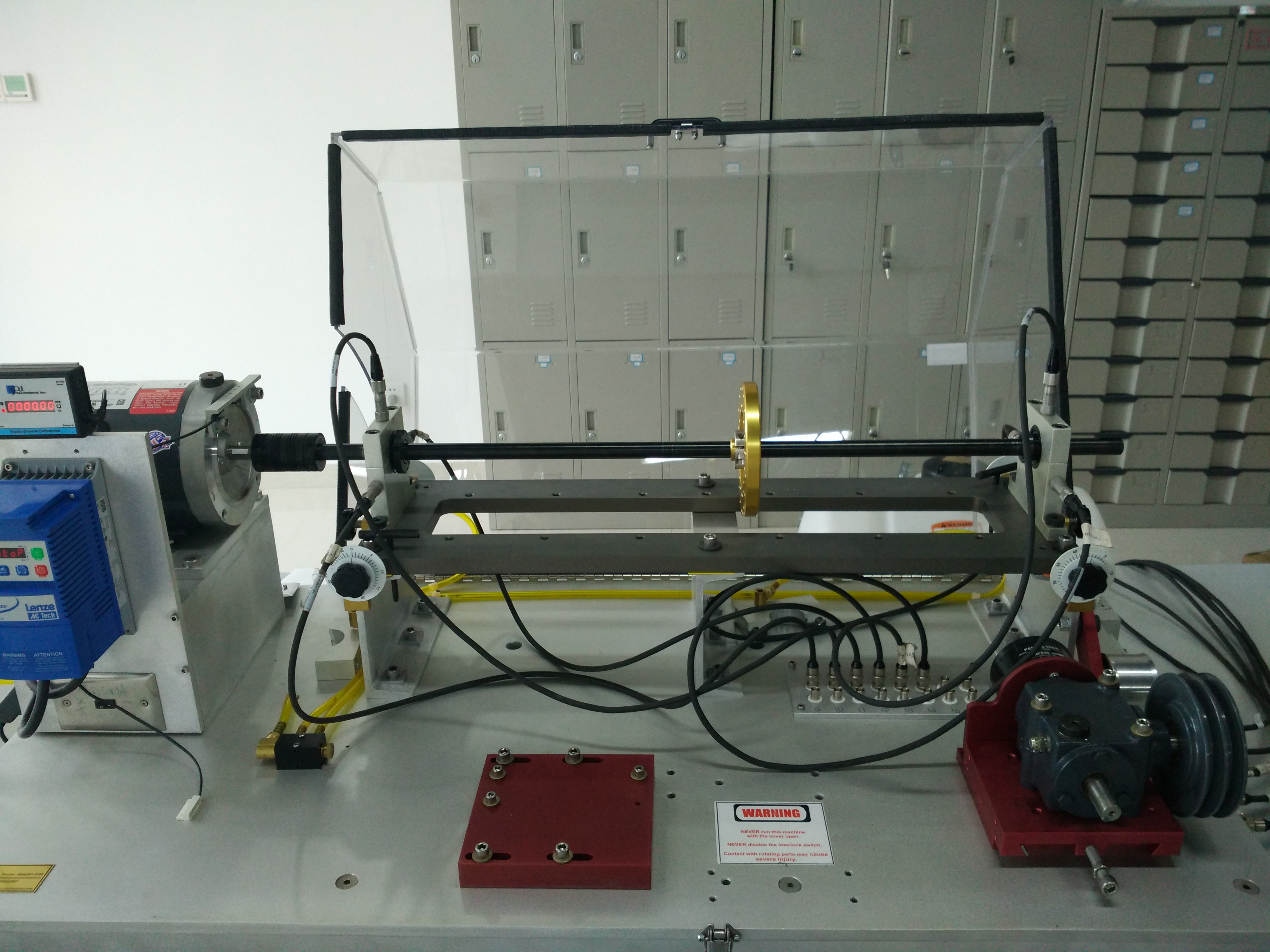}
  \caption{Machinery fault simulator experimental setup}
\label{Fig_lab_test_bed}
\end{figure*}

The database B used here is obtained from the accelerometers of the machinery fault simulator (Figure \ref{Fig_lab_test_bed}) at a sampling frequency of 20kHz from prof. Li lab. A well-balanced mass rotor is installed in the middle of a steel shaft which is supported by bearing housings with two rolling bearings. This simulator is driven by a 3 hp ac motor and several ICP accelerometers are mounted on the bearing housings. The speed of simulator is adjusted by the inverter and there are four speeds, 960rpm, 1080rpm, 1200rpm and 1320rpm. By replacing the bearing in the left bearing housing with the fault bearing, inner-race faults (IF), outer-race faults (OF) and ball fault (BF) with fault diameter being 0.75in are introduced into the machinery fault simulator. Finally, the vibration signals are collected by the ICP accelerometers on the top of the right bearing housing. The vibration signals of normal bearings (NO) under different load conditions were also gathered.

Four health conditions data (IF, OF, BF, NO) under one speed compose one dataset. There are four datasets totally (the names are in Table \ref{Table_lab_test_set}). So, there are 12 domain adaptation diagnosis problems under varied working condition totally. There are also 100 signals for each health condition and each sample contain 12000 data points. For convenience of calculations, we implemented FFT on each signal to get 8193 Fourier amplitudes, which form a sample. So each dataset contains 400 samples. In each test, one dataset is selected as the target domain, and the other three datasets are chosen as the source domains respectively. Each test is conducted twenty times, and  take the average accuracy as the final accuracy.

In this case, we will also compare the five methods, Baseline 1, Baseline 2, SVM NA, NN SA and SVM SA.

\begin{table}[htbp]
\begin{center}
  \caption{Description of the experiment datasets on database B}
  \label{Table_lab_test_set}
  \begin{tabular}{lccccl}
    \hline
    No. & Datasets & Speed & Fault type & Fault size & Sample size\\
    \hline
    1 & L0 & 960 rpm & IF,BF,OF,NO & $3/4^{''}$ & 400 \\
    2 & L1 & 1080 rpm & IF,BF,OF,NO & $3/4^{''}$ & 400 \\
    3 & L2 & 1200 rpm & IF,BF,OF,NO & $3/4^{''}$ & 400 \\
    3 & L3 & 1320 rpm & IF,BF,OF,NO & $3/4^{''}$ & 400 \\
    \hline
  \end{tabular}
  \end{center}
\end{table}

\subsubsection{Diagnosis results of the proposed method}

\begin{figure*}[!ht]
\centering
  \subfloat[]{\includegraphics[width=0.5\textwidth]{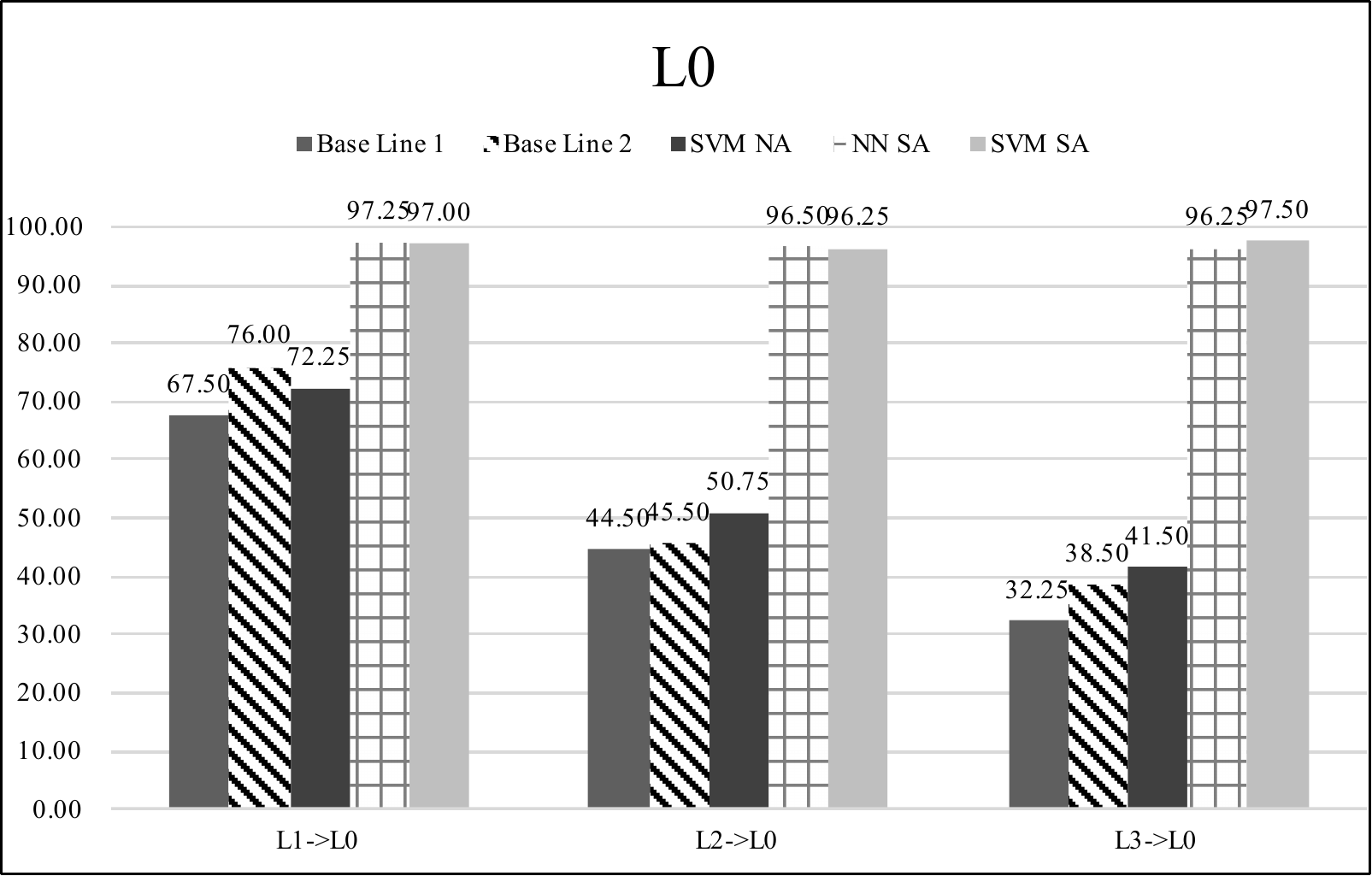}
  \label{fig_0_all_lab}}
  \subfloat[]{\includegraphics[width=0.5\textwidth]{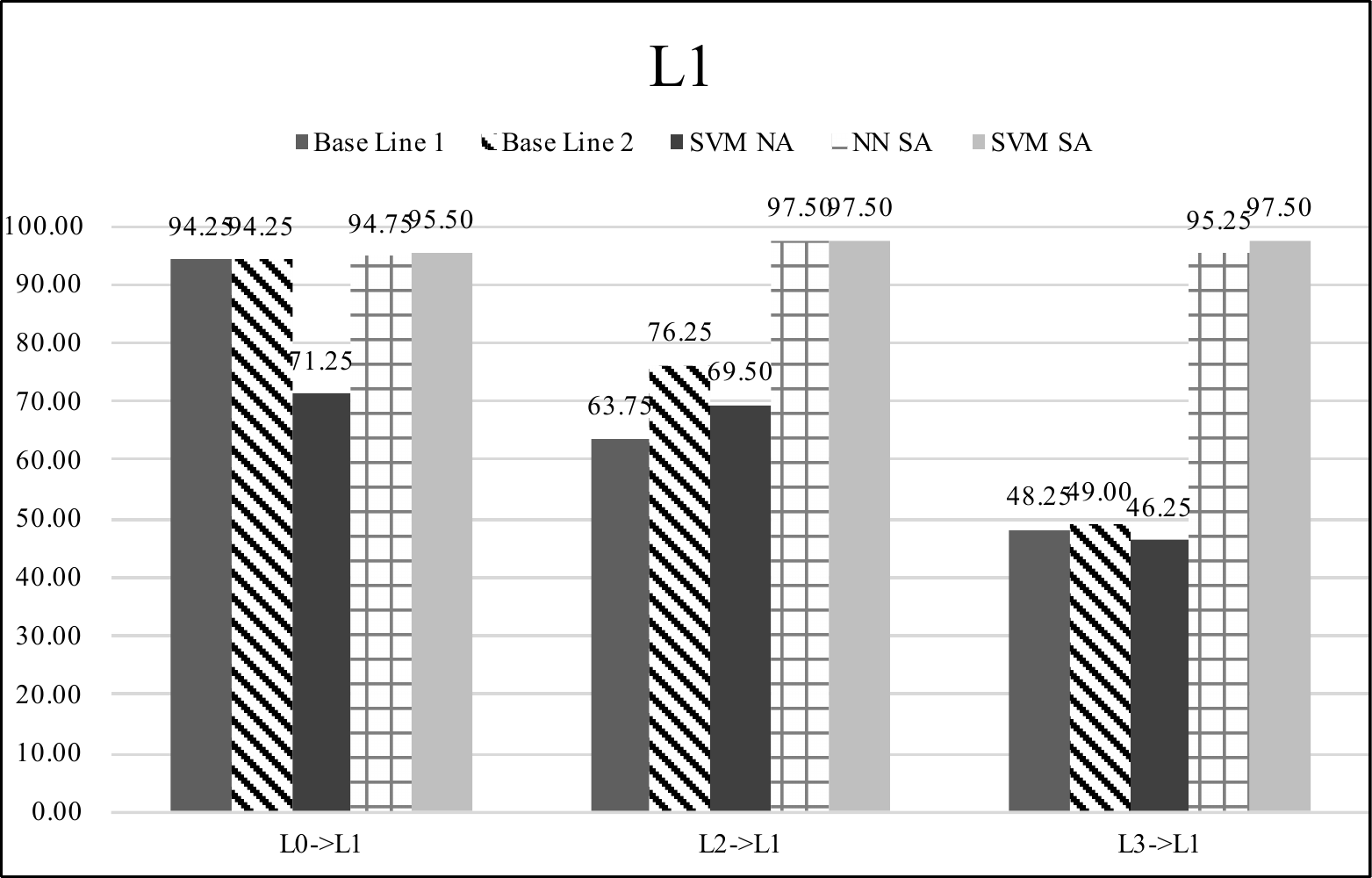}
  \label{fig_1_all_lab}}
  \hfil
  \subfloat[]{\includegraphics[width=0.5\textwidth]{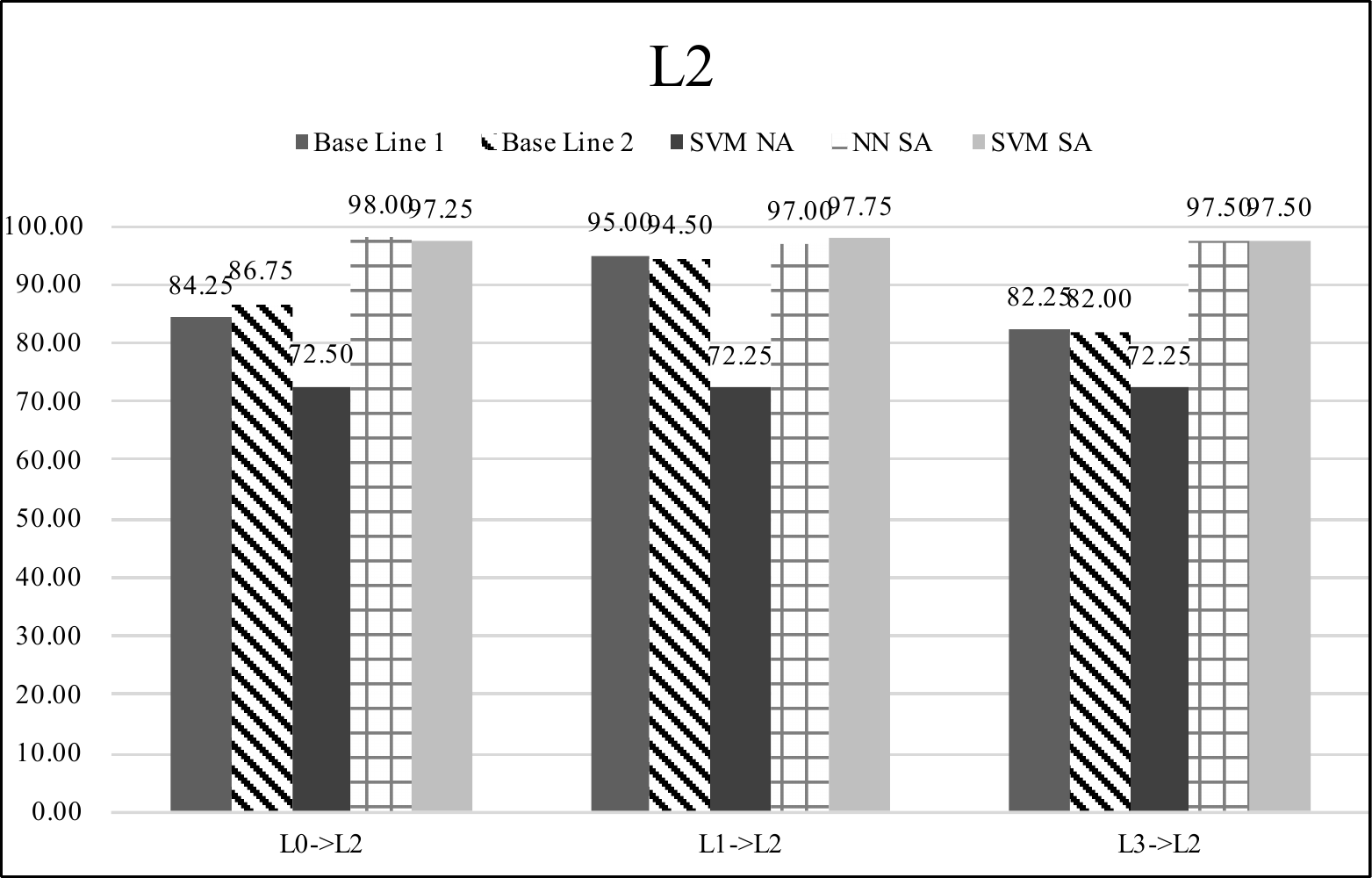}
  \label{fig_2_all_lab}}
  \subfloat[]{\includegraphics[width=0.5\textwidth]{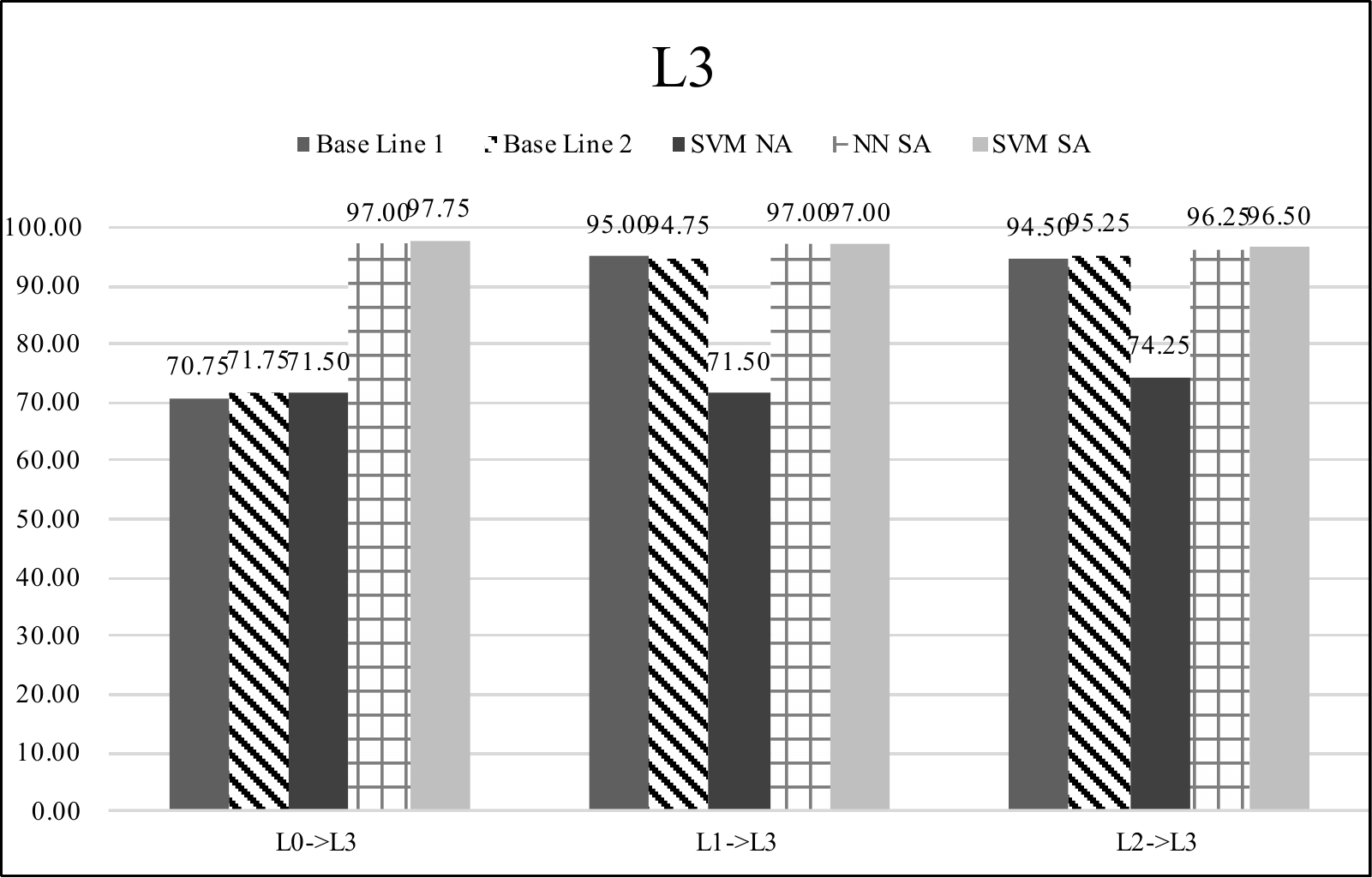}
  \label{fig_3_all_lab}}
\caption{The results of five methods based on database B}
\label{Fig_lab_all_load}
\end{figure*}

The diagnosis results of five methods are shown in Figure \ref{Fig_lab_all_load}. From this figure, we can find that the speed difference between training domain and testing domain is larger, the performances of methods with no adaptation, Baseline 1, Baseline 2 and SVM NA are poorer. It indicates that the difference of speed is greater, the distribution difference is greater. For example, in Figure \ref{fig_0_all_lab}, the testing domain is L0 (the speed is 960rpm), the training domains are L1 (the speed is 1080rpm), L2 (the speed is 1200rpm) and L3 (the speed is 1320rpm) respectively. The sort of the accuracies of methods with no adaptation in different training domains is $L1 > L2 >L3$. This phenomenon represents that this database is reasonable, to some degree because it conforms to the actual situation. It's also obvious that the performances of methods, NN SA and SVM SA, are superior to the other three methods, whatever the training domain and testing domain are. Unfortunately, the accuracies of the proposed two methods can't get $100\%$ like in Case 1. That is because the database B is not as ideal as database A. In general, the proposed two methods are more domain invariant than the traditional methods and the proposed methods also can be applied into this database.

\subsection{Discussion}
(1) This work present a new idea that use domain adaptation to conduct bearing fault diagnosis under varying working conditions. In ref. \cite{Li2016}, Li et al. proposed spectrum images method which applied two-dimensional principal component analysis (2DPCA) into the dimension reduction of the spectrum images of vibration signals, and the overall high accuracies were obtained. Unfortunately, there are still several instances having lower accuracies. To solve this problem, we apply unsupervised DA with SA into this field and the more data information from testing domain were used to train the classifier. Finally the accuracies almost all achieve $100\%$.

(2) The results of two diagnosis cases indicate that the proposed methods are able to effectively classify mechanical health conditions under varying working conditions and our methods are more domain adaptive. Actually, it's not fair to compare the accuracies of this method with method in Ref. \cite{Li2016}, because the method in Ref. \cite{Li2016} did't used the information from testing data. However, the proposed methods can be applied into more working conditions than the method in Ref. \cite{Li2016}. For example, in database A provided by Case Western Reserve University, the method in Ref. \cite{Li2016} just can conduct fault diagnosis under working conditions (Load0, Load1, Load2 and Load3) with the fault size being 0.014in and 0.021in. The method proposed in this paper can classify the faults under varying working conditions (Load0, Load1, Load2 and Load3) with the fault size being 0.007in, 0.014in, 0.021in and 0.028 in. Especially, the accuracies of the proposed method, SVM NA are all $100\%$, and the accuracies of the proposed method, NN NA are close to $100\%$.

(3)	In this paper, the method proposed belong to a linear method. The subspaces used to align are generated by PCA, which is a linear algorithm, and the alignment method is also a linear method. The classifiers are the general SVM and general NN. So, the complexity of the data this method can solve is limited. We will solve this problem in the future work.

\section{Conclusion}
\label{sec:conclusion}

When the working condition of rolling bearings varies, many traditional fault diagnosis models fail due to the phenomenon that the labeled training data and unlabeled testing data are drawn from the different distribution. Focusing on the fault diagnosis problem under varying working conditions, this paper presents a novel diagnosis strategy based on unsupervised DA using SA, which takes full advantage of large amounts of previously labeled training data and avoids annotating huge amount of new training data to rebuild model by spending lots of human efforts and time. In this method, we generate data space with simple FFT and create feature subspace by PCA for training data and testing data firstly. Then, a linear mapping is learnt to align the training subspace with the testing subspace under varying working conditions by unsupervised DA using SA. According to the visualization of data distributions, the same class of different domains does be pushed close to each other. Finally, a common linear classifier is trained based on the aligned training data, such as k-nearest-neighbor and SVM with precomputed kernel matrix, to predict the target values of the testing data. In order to fully benchmark the proposed method under varying working conditions, including different loads and speeds, we set up 60 domain adaptation diagnosis problems under varying working condition in Case Western Reserve benchmark data and 12 domain adaptation diagnosis problems under varying working condition in our new data. Different from Case Western Reserve benchmark data, our new collected experimental data focuses more on testing the influence of the speed-changing on the fault diagnosis algorithm. Experimental results show that the proposed method is more domain invariant than the traditional methods. After processed by unsupervised DA with SA, the distributions of training data and testing data are very close. The linear classifier trained on the aligned training data can be used to classify the testing data. Furthermore, the proposed method can effectively distinguish not only bearing faults categories but also fault severities.

\section*{Acknowledgment}

The work was supported by National Natural Science Foundation of China (51475455), Natural Science Foundation of Jiangsu (BK20160276), the Fundamental Research Funds for the Central Universities (2014Y05), and the project funded by the Priority Academic Program Development of Jiangsu Higher Education Institutions (PAPD).

\newpage

\section*{References}

\bibliography{secondpaper}

\end{document}